\documentclass[twocolumn]{aastex701}

\begin{document}

\title{Detection of a Millisecond Periodicity in BATSE Short Gamma-Ray Bursts}

\author[orcid=0009-0009-2083-1999]{Run-Chao Chen}
\affiliation{School of Astronomy and Space Science, Nanjing University, Nanjing 210093, China}
\affiliation{Key Laboratory of Modern Astronomy and Astrophysics (Nanjing University), Ministry of Education, China}
\email[show]{chrczxx@smail.nju.edu.cn}

\author[0000-0002-5485-5042]{Jun Yang}
\affiliation{Institute for Astrophysics, School of Physics, Zhengzhou University, Zhengzhou 450001, China}
\email[hide]{jyang@smail.nju.edu.cn}

\author[0000-0002-5596-5059]{Yi-Han Iris Yin}
\affiliation{Department of Physics, The University of Hong Kong, Pokfulam Road, Hong Kong, China}
\affiliation{The Hong Kong Institute for Astronomy and Astrophysics, The University of Hong Kong, Hong Kong, China}
\email[hide]{iris.yh.yin@connect.hku.hk} 

\author[0000-0003-4111-5958]{Bin-Bin Zhang}
\affiliation{School of Astronomy and Space Science, Nanjing University, Nanjing 210093, China}
\affiliation{Key Laboratory of Modern Astronomy and Astrophysics (Nanjing University), Ministry of Education, China}
\email[show]{bbzhang@nju.edu.cn}

\begin{abstract}
Coherent oscillations at kilohertz frequencies have recently been detected in a small number of gamma-ray bursts (GRBs), suggesting quasi-periodic dynamics in their central engines. A prominent example is GRB~230307A, which exhibited a brief, highly coherent, energy-dependent periodic signal interpreted as the possible spin signature of a nascent millisecond magnetar formed after a compact binary merger. Motivated by these developments, we conducted a comprehensive search for similar signals, accounting for both temporal and spectral dependencies, in 532 short GRBs with time-tagged event data recorded by the Burst and Transient Source Experiment (BATSE) onboard the \textit{Compton Gamma-Ray Observatory}. Within this sample, we identify a single statistically significant case: GRB~960616 (BATSE trigger~5502), in which the $\sim$30~ms main emission episode is coherently modulated at 1100~Hz, with the strongest modulation above 320~keV and a fractional amplitude of $\sim$47\%.
{Assuming the presence of a coherent periodic modulation, we use data-driven Monte Carlo simulations to place an upper limit of $\sim$8\% on the fraction of the total radiated energy that can be modulated by the QPO. This event, exhibiting a periodicity at $\sim$0.91~ms, further supports the possibility that millisecond periodicities can arise during GRBs in merger-driven scenarios.}
\end{abstract}


\section{Introduction}\label{sec:intro}
The progenitor systems of gamma-ray bursts (GRBs), broadly categorized into massive star type \citep{1993ApJ...405..273W, 1999ApJ...524..262M} and compact binary type \citep{1989Natur.340..126E, 1991AcA....41..257P, 1992ApJ...395L..83N, 2014ARA&A..52...43B}, undergo catastrophic events that leave behind a central engine powering a collimated relativistic jet responsible for the prompt gamma-ray emission. This engine is generally thought to be either a hyper-accreting black hole or a rapidly rotating, highly magnetized neutron star (millisecond magnetar) \citep{1992Natur.357..472U, 1992ApJ...392L...9D, 1992ApJ...395L..83N, 1993ApJ...405..273W, 1999ApJ...524..262M, 1998A&A...333L..87D, 2001ApJ...550..410M, 2006PhRvL..96c1102S, 2011MNRAS.413.2031M, 2014MNRAS.439.3916M}. While the general connections between GRBs and their progenitor systems have been well established through gravitational wave detections and multi-wavelength observations \citep{1998Natur.395..670G, 2003ApJ...591L..17S, 2006Natur.444.1010Z, 2009ApJ...703.1696Z, 2017ApJ...848L..12A, 2017PhRvL.119p1101A, 2020ApJ...897..154L}, the physical nature of the central engines formed in these catastrophic events remains uncertain \citep{2011CRPhy..12..206Z, 2015PhR...561....1K}.

Recent breakthroughs have provided new observational pathways toward probing the GRB central engine through high-frequency variability in the prompt emission. \citet{2023Natur.613..253C} reported a quasi-periodic oscillation (QPO) at $\sim$2.7~kHz in a short GRB, suggesting that a rapidly rotating hypermassive neutron star may have temporarily survived the merger and modulated the prompt emission. However, the lack of follow-up data precluded a detailed physical interpretation of the signal. More recently, \citet{2025NatAs...9.1701C} identified a brief but stable 909~Hz periodic modulation in the merger-type GRB~230307A. Owing to the exceptionally bright prompt emission and extensive multi-wavelength coverage of this event \citep{2024Natur.626..737L, 2024Natur.626..742Y, 2025NSRev..12E.401S}, it was possible to establish the first physically consistent association between a periodic GRB signal and its central engine, interpreted as a nascent millisecond magnetar \citep{2025NatAs...9.1701C}. This finding also provided a plausible mechanism for generating such periodicities during the magnetic dissipation stage of a relativistic outflow \citep{2025NatAs...9.1701C}.

Motivated by these recent discoveries, we revisit the archival data from the Burst and Transient Source Experiment (BATSE) onboard the \textit{Compton Gamma-Ray Observatory} (\textit{CGRO}). In particular, we aim to investigate whether signatures similar to those observed in GRB~230307A, potentially indicative of a millisecond magnetar central engine, can also be identified in canonical merger-type bursts with durations below 2~s \citep{1993ApJ...413L.101K, 2014ARA&A..52...43B}. The eight Large Area Detectors (LADs) of BATSE offer nearly all-sky coverage, each with an effective area of $\sim$2000~cm$^2$ and a timing resolution of 2~$\mu$s, making them exceptionally suited for detecting millisecond-scale periodic or quasi-periodic modulations in GRBs. In this work, we perform a systematic search for high-frequency periodic signals in the Time-Tagged Event (TTE) data of 532 short GRBs detected by BATSE \citep{1999ApJS..122..465P}. Building on the methodological framework established by \citet{2025NatAs...9.1701C}, our analysis explicitly accounts for both the time and energy dependence of potential periodic modulations, ensuring sensitivity to transient and energy-selective oscillatory behavior.

This paper is organized as follows. 
Section~\ref{sec:instrument} briefly summarizes the BATSE instrumentation and the characteristics of its TTE data. 
Section~\ref{sec:search} describes the data reduction procedures and the statistical methodology, along with the results of the periodicity search.
Section~\ref{sec:5502} presents a comprehensive validation and time-domain analysis of the only significant periodic candidate identified in this study, GRB~960616 (trigger5502), integrating consistency checks, timing analyses, and data-driven modeling to assess the robustness and physical significance of the signal.
Section~\ref{sec:sum} summarizes the major findings and discusses their implications for the nature of the central engine in short GRBs.

\section{Instrumentation and Data Types}\label{sec:instrument}
\subsection{Instrument Overview}
The BATSE instrument onboard the \textit{CGRO} consisted of eight uncollimated detector modules mounted at the corners of the spacecraft, providing nearly full-sky coverage for detecting high-energy transients\footnote{The following summary is based on the official BATSE instrument documentation available at \url{https://heasarc.gsfc.nasa.gov/docs/cgro/batse/}.}. Each module equipped two NaI(Tl) scintillation detectors: a Large Area Detector (LAD) and a Spectroscopy Detector (SD). The LAD, optimized for burst detection and directional response, was a 50.8~cm diameter, 1.27~cm thick NaI(Tl) crystal viewed by photomultiplier tubes (PMTs), giving an effective area of $\sim$2025~cm$^2$. The SD, measuring 12.7~cm in diameter and 7.6~cm in thickness, operated at a higher PMT gain to extend the measurable energy range from $\sim$10~keV up to $>$100~MeV with improved spectral resolution.

Scintillation pulses from the detectors are processed by a gated baseline restoration circuit and fed into both a high-speed four-channel discriminator system and a slower pulse-height analysis chain. The nominal equivalent energies of the upper three discriminators for the LADs are 60, 110, and 325~keV, while the programmable lower-level threshold is typically set near 20~keV. For the SDs, two of the fast discriminators are placed above the energy range analyzed by the pulse-height system to extend the high-energy response. Pulse-height converters form 128-channel spectra from the LADs and 256-channel spectra from the SDs, which are further mapped by the Central Electronics Unit (CEU) into 16 coarse energy channels using programmable lookup tables. This design allows flexible trade-offs between time and energy resolution in different telemetry data types\footnote{Detailed descriptions of the BATSE's archival data types are available at \url{https://astro-gdt-cgro.readthedocs.io/en/latest/missions/cgro/batse/gallery.html}.}. 

\subsection{Time-Tagged Event Data}
Among the several data products generated onboard, the TTE mode provides the highest temporal resolution. In this mode, the CEU records the arrival time, discriminator channel, and detector number of each individual photon detected by the LADs. Each triggered burst fills a dedicated buffer capable of storing up to 32,768 photon events with a nominal 2~$\mu$s timing resolution. The memory operates as a circular buffer: upon a burst trigger, photon accumulation continues until three-fourths of the buffer is filled, while the remaining quarter preserves pre-trigger events from all eight LADs, thereby providing both pre- and post-trigger coverage.

Each TTE event thus carries photon-by-photon information including arrival time, discriminator index, and detector number. The discriminator channels correspond approximately to photon energies within 25--60~keV, 60--110~keV, 110--325~keV, and $>$325~keV, respectively, providing coarse energy tagging that is sufficient for studies of time-dependent hardness and variability. Compared with other BATSE telemetry modes, the TTE mode uniquely preserves the unbinned photon stream, enabling sub-millisecond reconstruction of burst light curves and sensitive searches for coherent or quasi-periodic oscillations.

\section{Search for periodicities in TTE data of BATSE short GRBs}\label{sec:search}
\subsection{Data reduction}\label{subsec:reduction}
In this work, we extracted and analyzed TTE data for 532 short GRBs from the BATSE archive, covering triggers recorded between May 1991 and May 2000. Given that periodic signals in GRBs can appear briefly and exhibit strong energy dependence \citep{2025NatAs...9.1701C}, we performed time- and energy-resolved analyses for each burst. Specifically, we first extracted photon arrival times, energy-channel indices, and detector identifiers from the \texttt{ASCII}-format TTE files\footnote{The \texttt{ASCII}-format TTE files for each burst are publicly available at \url{https://heasarc.gsfc.nasa.gov/FTP/compton/data/batse/ascii_data/batse_tte/}. Each file records photon events only from the detectors that were triggered during the burst. The trigger configuration of BATSE was determined solely by the observed photon flux and was independent of pre-screened source or Earth angle constraints.}. The analysis was then carried out separately for each triggered detector and across the four standard BATSE energy discriminator channels (nominally 25--60~keV, 60--110~keV, 110--325~keV, and $>$325~keV).

To construct time windows suitable for high-resolution timing analysis, we first applied the Bayesian blocks algorithm \citep{2013ApJ...764..167S} to the merged photon list of each burst, combining data from all triggered detectors and energy channels to obtain a global segmentation of statistically homogeneous temporal intervals. The resulting block structure was then adopted as the reference time grid for constructing detector- and channel-resolved time windows. Within each Bayesian block, we generated a sequence of overlapping 100~ms sliding windows, with each window advanced by a fixed 50~ms step relative to the previous one. A protection condition was applied to ensure that the end time of any window never exceeded the end time of the available TTE record, thereby guaranteeing that the final segment remained entirely within the valid data range. The detailed segmentation procedure is described in Appendix~\ref{sec:a1} and illustrated in Figure~\ref{fig_a1}. This adaptive yet continuous segmentation procedure provides adequate temporal sampling for detecting transient periodic modulations while maintaining complete coverage for each burst.

The TTE data inherently contain both source and background intervals. However, we did not restrict the analysis time range to the source intervals, but deliberately retained this full time coverage for two reasons. First, some bursts may exhibit extended or weak extended emission components (e.g., the case presented in Appendix~\ref{sec:a1}), which could be missed if the analysis were restricted to a narrow time range. Second, the periodicity search is conducted under a null prior hypothesis that the observed variations contain no intrinsic (quasi-)periodic components, and all apparent peaks in the power spectrum are assumed to arise from stochastic noise processes. Including background intervals therefore does not bias the analysis. Even if spurious periodic signals were to arise within background-dominated regions, our method allows precise temporal localization of such detections, enabling retrospective verification of their association with the main emission episodes.

This reduction and segmentation scheme ensures that all bursts are processed in a uniform and reproducible manner, while preserving potential temporal structures of interest for subsequent statistical analyses. In total, the analysis yielded a sample of 186,327 overlapped 100~ms time-segmented event datasets from 532 short GRBs, covering four distinct energy channels and multiple triggered detectors, each serving as an individual trial for periodicity testing.

\subsection{Methodology}\label{subsec:methodology}
After segmenting the event data in both time and energy, we conducted a blind search for coherent periodicity using the Rayleigh test \citep{10.1111/j.2517-6161.1975.tb01550.x}. The Rayleigh power at a trial frequency $f$ is defined as
\begin{equation}
R(f) = \frac{1}{N}\left[\left(\sum^N_{j=1}\cos(2\pi f t_j)\right)^2 + \left(\sum^N_{j=1}\sin(2\pi f t_j)\right)^2\right],
\end{equation}
where $N$ is the number of photons arriving at time $t_j$ relative to the reference time $T_0$\footnote{For each trigger, the time series $t_j$ is constructed from the recorded photon arrival times in the \texttt{ASCII} file, expressed in units of $1\times10^{-6}$~s, with $T_0 \equiv 0$~s.}. We scanned trial frequencies between 500 and 2,500~Hz with a resolution of 1~Hz. For each segment, the maximum Rayleigh power across all trial frequencies was recorded as the candidate power $R_x$, with the corresponding frequency $f_x$ identified as the candidate frequency.

Under the null hypothesis that no intrinsic (quasi-)periodicity exists, all apparent peaks in the periodogram are assumed to originate from stochastic noise fluctuations. According to extreme-value theory \citep{1928PCPS...24..180F}, the maximal Rayleigh powers $R_x$ obtained from each segment, as well as any sufficiently large subset of $\{R_x\}$, follow the same Gumbel (type-I extreme-value) distribution characterized by the cumulative form \citep{mood1950introduction}:
\begin{equation}
G(R_x) = e^{-e^{-(R_x-\mu)/\beta}},
\label{eq:cdf}
\end{equation}
where $\mu$ and $\beta$ are the location and scale parameters, respectively. These parameters can be estimated by the method of moments:
\begin{equation}
{\rm E}[R_x] = \mu + \gamma\beta, \quad {\rm V}[R_x] = \frac{\pi^2}{6}\beta^2,
\label{eq:parms}
\end{equation}
with $\gamma\simeq0.577216$ being Euler's constant.

Following the approach of \citet{2025NatAs...9.1701C}, the reliability of a potential periodic signal was assessed through two independent statistical steps. (1)~A genuine periodic modulation must produce an extreme value that exceeds the expectation of a noise-induced ensemble. This was evaluated by conducting independent background trials to empirically determine the noise threshold. (2)~Even if an individual subset yields an extreme value, the global significance of the signal depends on how its power compares to the statistical expectation across all trial frequencies. 

Under the null hypothesis that the noise process has an equal chance of producing a false detection at any frequency, the single-trial probability is given by
\begin{equation}
p = \frac{N_{\rm freq}}{N_{\rm segment}},
\end{equation}
where $N_{\rm freq}$ is the number of trial frequencies, and $N_{\rm segment}$ is the number of temporal segments\footnote{Although adjacent segments are not strictly independent due to temporal overlap, the condition $N_{\rm segment} \gg N_{\rm freq}$ ensures that the effective false-alarm probability remains conservatively estimated.}. Hence, the probability that none of the $N_{{\rm cand},i}$ independent candidates at a given frequency $f_i$ exceed the noise expectation is
\begin{equation}
{\rm Prob}_{{\rm noise}, i} = (1 - p)^{N_{{\rm cand}, i}}.
\end{equation}
This probability defines the expected maximal Rayleigh power arising purely from stochastic fluctuations at that frequency:
\begin{equation}
R_{{\rm noise}, i} = \mu_{f_i} - \beta_{f_i}\ln\!\left[-\ln\!\left(1 - {\rm Prob}_{{\rm noise}, i}\right)\right],
\end{equation}
where $\mu_{f_i}$ and $\beta_{f_i}$ are the Gumbel location and scale parameters, respectively, obtained by fitting the local ensemble of candidate powers at $f_i$.

The overall distribution of the expected maximal noise powers $R_{{\rm noise}, i}$ across all trial frequencies is modeled using a cumulative distribution function $\mathcal{H}(R)$, which shares the same functional form as the global $G(R_x)$ defined in Equation~(\ref{eq:cdf}), but allows for refitted parameters. Specifically, this relation can be expressed as $\mathcal{H}(R) = G(R;\,\mu, \beta, \xi)$, where $\mu$, $\beta$, and $\xi$ are the location, scale, and shape parameters, respectively, estimated from the ensemble of candidate powers across all trial frequencies. This formulation allows $\mathcal{H}(R)$ to belong to the Generalized Extreme Value family, while the specific case of $\xi = 0$ naturally reduces to the Gumbel form in Equation~(\ref{eq:cdf}).

The single-trial tail probability for frequency $f_i$ is therefore defined as $p_i = 1 - \mathcal{H}(R_{{\rm noise}, i})$. Assuming independent trials across the $N_{\rm freq}$ sampled frequencies, the corresponding false-alarm probability (FAP) is given by
\begin{equation}
{\rm FAP}_{f_i} = 1 - \left[\mathcal{H}(R_{{\rm noise}, i})\right]^{N_{\rm freq}},
\end{equation}
which represents the probability that at least one of the $N_{\rm freq}$ frequency trials yields a noise fluctuation with power equal to or exceeding the observed $R_{{\rm noise}, i}$. If the independence assumption is not fully satisfied, a conservative Bonferroni upper bound can be adopted as 
${\rm FAP}_{f_i} \le N_{\rm freq}\times p_i$.

\subsection{Results for the search of periodicity}
Given that the 186,327 segments include short GRBs with widely varying signal-to-noise ratios and substantial background intervals, and considering the large effective area of the BATSE-LADs that increases the likelihood of recording extreme particle events, we first performed a preliminary screening to exclude extreme high-power outliers arising from red-noise contamination. During this process, we found one anomalous case with trigger number 6634 (corresponding to GRB~980310B). As shown in Figure~\ref{fig_a2}, this burst exhibits an extremely short ($\sim$2~ms) and intense spike in its light curve, producing pronounced red-noise excess extending up to 2000~Hz\footnote{This burst also appears as a typical false-positive example in Extended Data Figure~2 of \citet{2023Natur.613..253C}.}. Because this spike appears consistently across multiple triggered detectors and energy channels, it generates numerous spurious high-power outliers. To avoid biasing the statistical distribution, we excluded all data segments from this GRB. 

\begin{figure}
\centering
\includegraphics[width=0.9\linewidth]{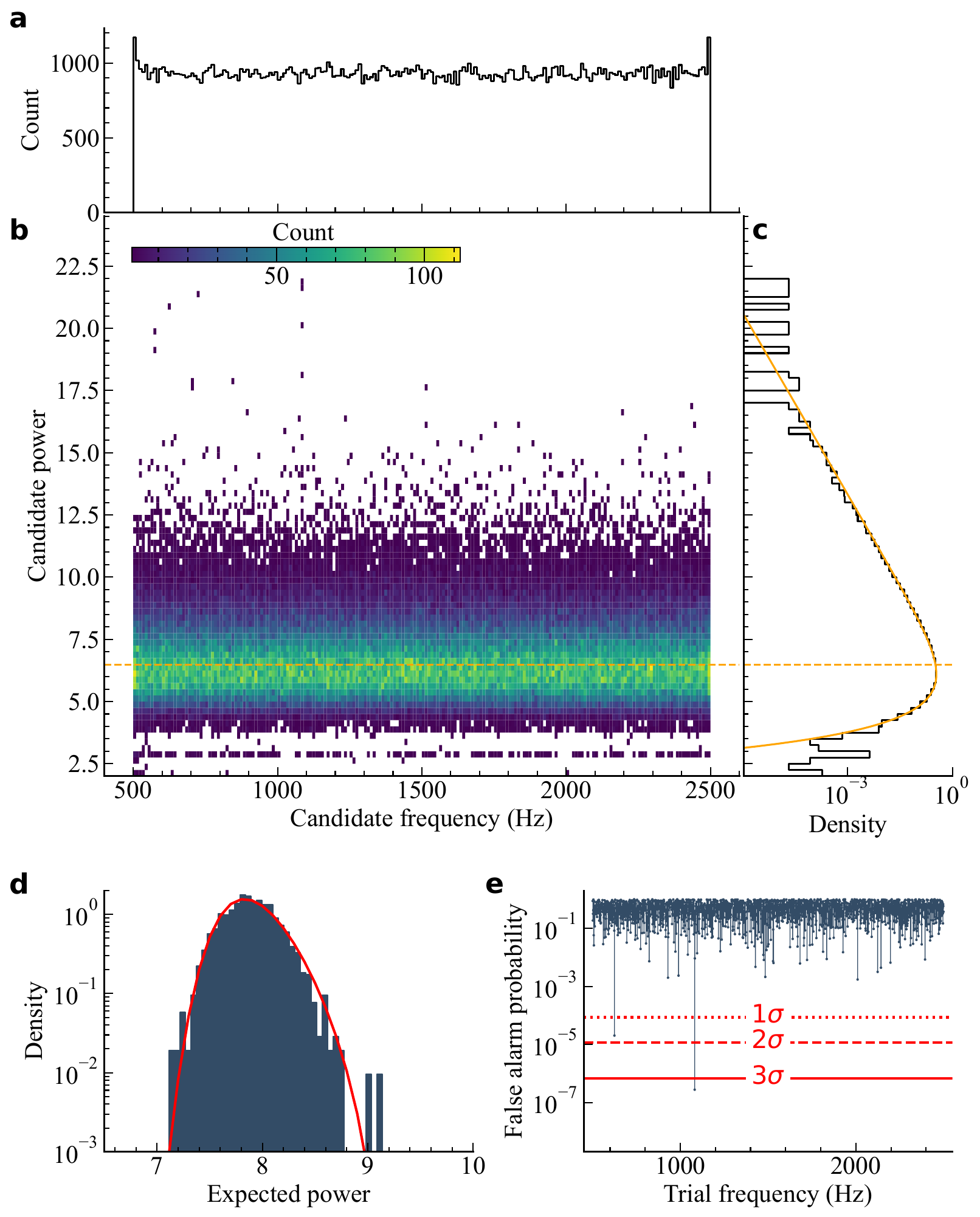}
\caption{\noindent\textbf{Results of the periodicity search in BATSE TTE data.}
\textbf{a--c:} Marginal and joint distributions of 186,127 candidate frequencies ($f_x$) and Rayleigh powers ($R_x$) derived from 100-ms segments extracted from different LADs and energy channels of 531 short GRBs (excluding trigger~6634). The orange curve in \textbf{c} shows the probability density function (p.d.f.) of the ensemble of all $R_x$. Orange dashed lines in \textbf{b} and \textbf{c} mark the median threshold power $R_{50}$ defined by $G(R_{50})=0.5$. 
\textbf{d:} Distribution of expected powers derived from 94,766 segments with $R_x \ge R_{50}$ at trial frequencies between 500 and 2500~Hz. The red curve denotes the asymptotic p.d.f. obtained from a generalized extreme-value distribution ($\mathcal{H}(R)$) fit.
\textbf{e:} Single-trial FAPs as a function of trial frequencies. The red dashed lines indicate significance levels after accounting for the number of frequency trials. Only trigger~5502 shows a $>3\sigma$ excess at 1083~Hz, while trigger~6412 exhibits a weaker $\sim$1$\sigma$ excess at 625~Hz.}
\label{fig_cand}
\end{figure}

After the preliminary screening, the final ensemble used for the statistical analysis, as shown in Figure~\ref{fig_cand}\textbf{a--c}, contains 186,127 candidate powers $R_x$ and their corresponding frequencies $f_x$. As noted in Section~\ref{subsec:reduction}, the dataset includes a substantial number of segments from background intervals across different triggers. As a result, it is not feasible to construct an equivalent set of background-only trials to empirically determine a noise threshold. Instead, we adopt a conservative statistical approach by approximating the null (noise) distribution directly from the global ensemble of all $R_x$, fitting a Gumbel probability density function in accordance with Equations~\ref{eq:cdf} and~\ref{eq:parms}. The resulting Gumbel distribution $G(R_x)$ characterizes the expected statistics of stochastic extremes. We define its median, $R_{50}$, such that $G(R_{50}) = 0.5$, as a reference threshold that separates the noise-dominated and signal-enhanced regimes for subsequent analysis. 

All candidates with $R_x > R_{50}$ were retained for further evaluation. This approach remains valid under the assumption that the ensemble of candidate powers is itself drawn from the same parent extreme-value distribution expected for noise. Figure~\ref{fig_cand}~\textbf{d--e} show the results of step~(2) described in Section~\ref{subsec:methodology}, where the relative significance of each candidate frequency is evaluated within the tail of the local Gumbel distribution.

During this procedure, two exceptional triggers were identified. The first, trigger~6412 (corresponding to GRB~971004-), exhibited a marginally significant feature at 625~Hz (Figure~\ref{fig_cand}\textbf{e}). This feature arises from segments that appear only in a single detector and energy channel, suggesting that it is unlikely to have a genuine physical origin. Moreover, as shown in Figure~\ref{fig_a2}, the apparent excess in the periodogram arises from a segment containing a bright, isolated spike that is temporally well separated from the main burst. Given its confinement to a single detector and low-energy channel, and lack of temporal coincidence with the main emission episode, we conclude that this feature is unlikely to represent a genuine periodic signal.

By contrast, trigger~5502 (corresponding to GRB~960616) exhibits a significant feature at 1083~Hz, with a trial-corrected false-alarm probability of ${\rm FAP} \approx 5.54\times10^{-4}$. When examining the trial frequencies within $1083\pm10$~Hz, this feature is found to arise from 15 overlapping segments spanning multiple detectors and energy channels, all of which temporally coincide with the main emission episode of a bright burst.
The significance of the peak and its recurrence across several consistent segments together indicate that the signal is consistent with a genuine periodic modulation. In the following section, we perform a detailed validation and characterization of this feature in GRB~960616.

\section{Validation of the periodicity in GRB~960616}\label{sec:5502}
GRB~960616 (trigger~5502) triggered BATSE at 1996-06-16T18:28:15.6~UTC. This burst lasted about 0.1~s and included a possible precursor occurring $\sim$10~ms before the main burst. In the following, we validate and characterize the properties of the possible periodic signal at around 1100~Hz using multiple analytical approaches.

\subsection{Consistency Check with Wavelet Analysis}
To examine the temporal evolution of the detected signal and verify its presence across different detectors, we applied the weighted wavelet Z-transform (WWZ) \citep{1996AJ....112.1709F} to generate time--frequency spectrograms for all triggered LADs. WWZ is a wavelet-like time--frequency analysis method designed for irregularly sampled data. Instead of computing the convolution of the signal with a kernel function as in traditional wavelet transforms, WWZ performs a localized, weighted least-squares fit of sinusoidal functions within a moving Gaussian window. This approach retains the temporal localization properties of the Morlet wavelet \citep{GrossmannMorlet1984} while yields a well-defined statistical measure, making it possible to directly compare the significance of oscillatory power across different time and frequency bins.

In this work, we calculate the WWZ power spectra using the \texttt{libwwz} package\footnote{An example of this \texttt{Python} package is available at \url{https://github.com/ISLA-UH/libwwz/blob/master/examples/example_wwz.py}.}. The event data within the time interval [0.06, 0.16]~s of GRB~960616 were binned into 0.1~ms bins, with each bin assigned a midpoint timestamp. The WWZ statistics were evaluated on a frequency grid from 500 to 2500~Hz with a 10~Hz step and a temporal resolution of 1~ms. The decay constant, which controls the balance between temporal and frequency resolution in the WWZ analysis, was set to $c = 1/(1100~\mathrm{Hz}\times0.1~\mathrm{s})^2$, consistent with the 0.1~s light curve duration and the expected periodicity near 1100~Hz.

\begin{figure}
\centering
\includegraphics[width=0.9\linewidth]{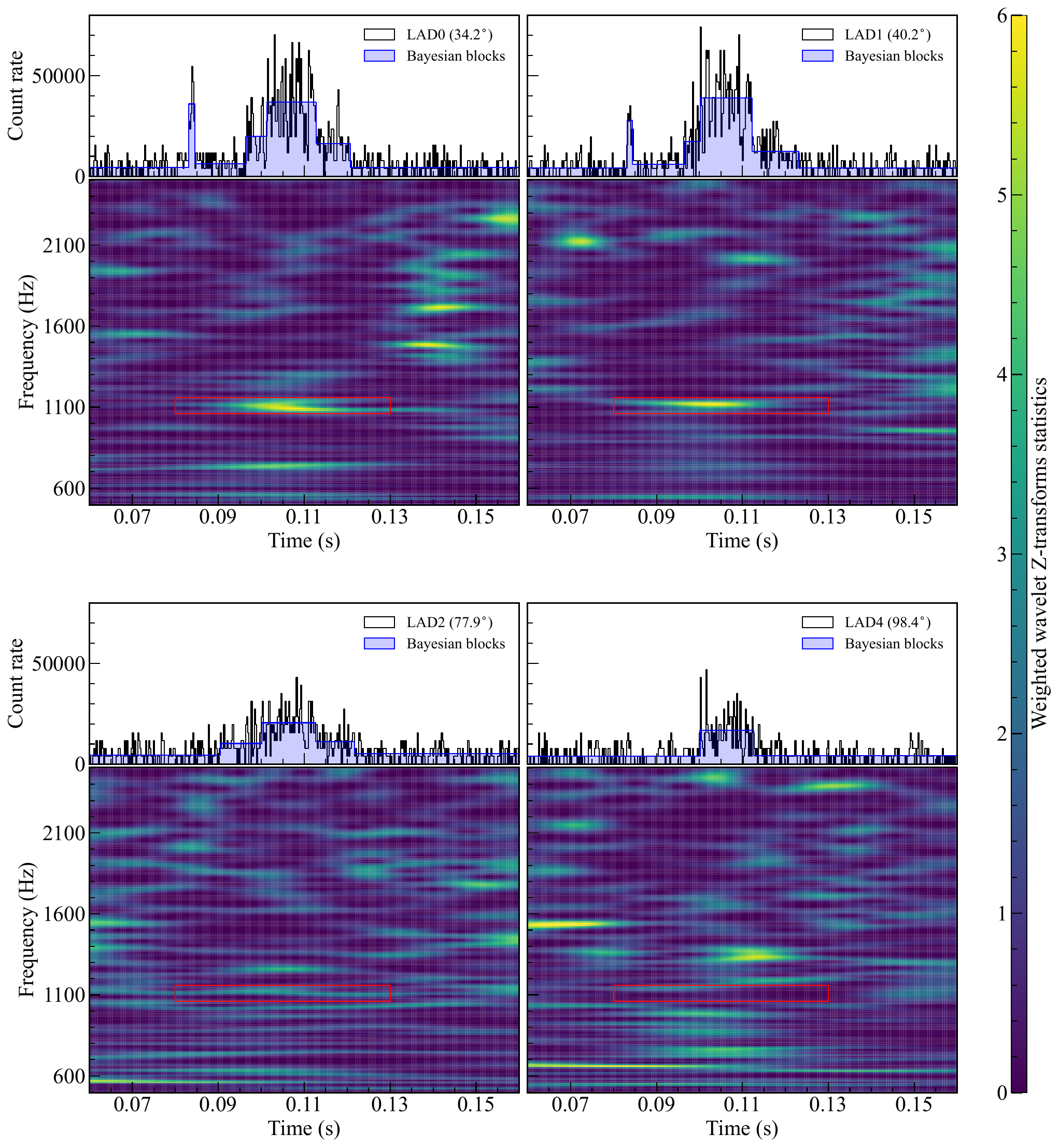}
\caption{\noindent\textbf{WWZ spectrograms of GRB~960616 (detector resolved).}
Light curves recorded by different LADs are shown at a temporal resolution of 256~$\mu$s, with the corresponding incident angles indicated. Red boxes highlight the $\sim$1100~Hz signal, which is clearly detected by LAD0 and LAD1 but not prominently observed in LAD2 and LAD4.}
\label{fig_wwz1}
\end{figure}

Figure~\ref{fig_wwz1} presents the WWZ spectrograms of GRB~960616 for the four individual triggered LADs, each combining photons from all four energy channels. A prominent oscillation at $\sim$1100~Hz is evident in LAD0 and LAD1, both of which had small incident angles relative to the burst direction. The oscillation persists throughout the burst duration, suggesting a coherent temporal modulation of the emission episode.

\begin{figure}
\centering
\includegraphics[width=0.9\linewidth]{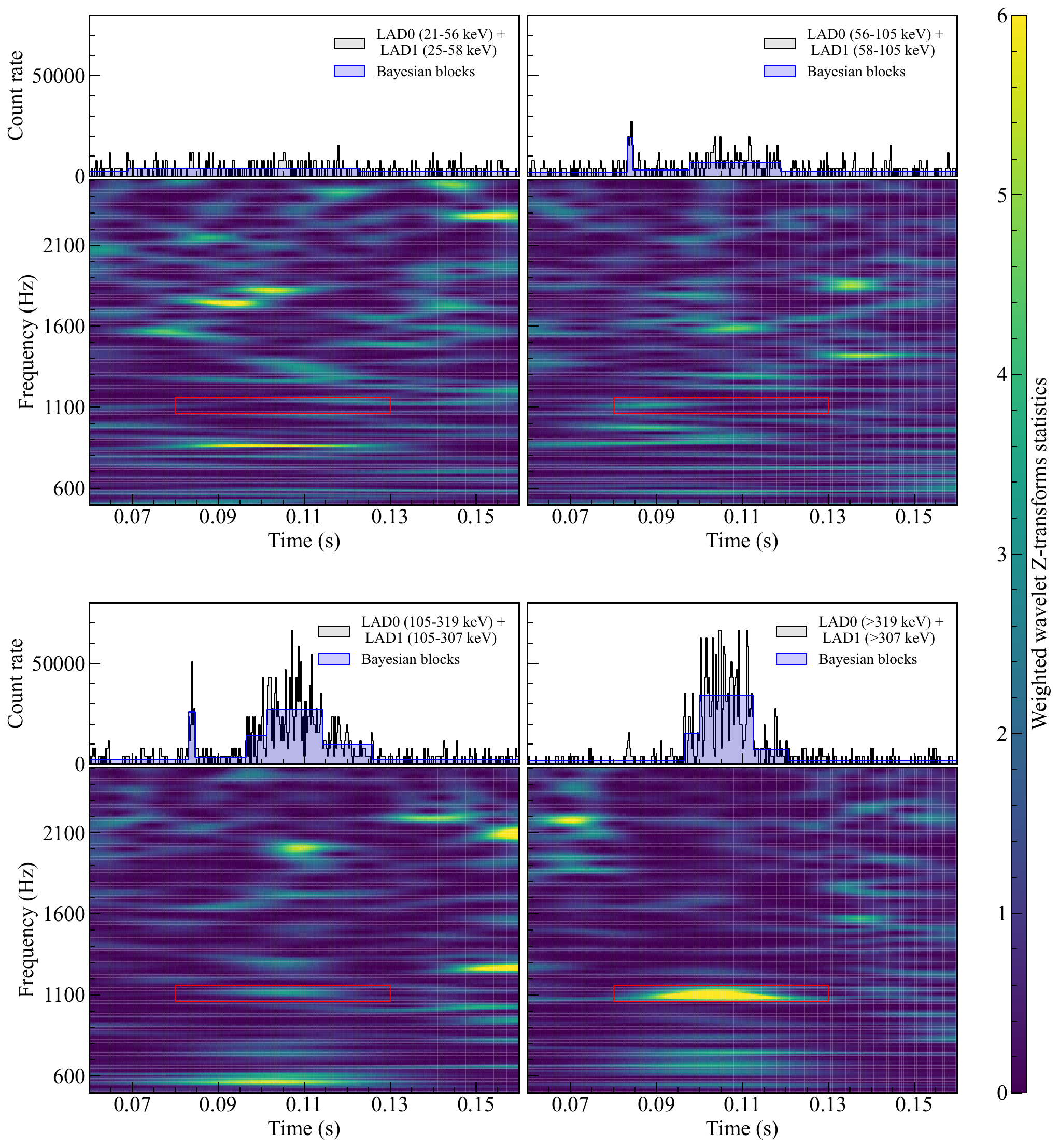}
\caption{
\noindent\textbf{WWZ spectrograms of GRB~960616 (energy-channel resolved).}
Light curves from LAD0 and LAD1 are combined and shown at a temporal resolution of 256~$\mu$s for the four BATSE energy channels, with the corresponding energy ranges indicated. Red boxes highlight the $\sim$1100~Hz signal, which is clearly detected in channel~4 and marginally visible in channel~3.}
\label{fig_wwz2}
\end{figure}

To further examine its energy dependence of the signal, we combined data from LAD0 and LAD1 to improve the signal-to-noise ratio and applied the same WWZ analysis to the four standard BATSE energy channels. As shown in Figure~\ref{fig_wwz2}, no significant oscillation is detected in channels~1 and~2, where the photon statistics are relatively poor. A marginal signal appears in channel~3, implying weak modulation at this frequency, while channel~4---the highest-energy band with the best signal-to-noise ratio---shows a clear and significant 1100~Hz modulation.

We also derived Bayesian block representations of the light curves in different energy bands and LADs. As shown in Figure~\ref{fig_wwz2}, the precursor emission is evidently softer than the main burst. Moreover, as seen in Figure~\ref{fig_wwz1}, the 1100~Hz oscillation becomes significant only during the main burst, indicating that this periodic modulation most likely originates from the main prompt emission episode rather than the precursor phase, which may arise from a distinct or less coherent emission process.

\subsection{Quasi-periodic nature of the 1.1~kHz signal}
Motivated by the time–frequency analysis presented in the previous subsection, which revealed a strong modulation near $\sim$1100~Hz with clear energy dependence, we further investigate this feature in the Fourier domain. Our aim is to assess whether the signal represents a broadened quasi-periodic feature or a coherent periodic oscillation, and to quantify its strength, characteristic frequency, and coherence relative to a no-QPO baseline.

Following the Bayesian methodology of \citet{2023Natur.613..253C}, we computed the power density spectrum (PDS) for GRB~960616 using a 0.131072~s interval starting at $t=0.085$~s, extracted from the combined LAD0 and LAD1 data in BATSE energy channel~4. This selection targets the time and energy range where the modulation is strongest (Figure~\ref{fig_wwz2}). The PDS is normalized following \citet{1975ApJS...29..285G}, so that the expectation value of the power is unity under pure Poisson noise.

The QPO component is modeled with a Lorentzian profile:
\begin{equation}
P_{\rm QPO}(f) \;=\; \frac{N}{\pi\,\Delta f_0}\,\frac{(\Delta f_0)^2}{(f-f_0)^2+(\Delta f_0)^2},
\label{eq:lor}
\end{equation}
where $f$ is the frequency, $f_0$ is the centroid frequency, $\Delta f_0$ is the half width at half maximum (HWHM), and $N$ is the integrated amplitude (so that $A \equiv N/[\pi \Delta f_0]$ gives the power density at $f_0$). The \emph{null} model allows for a constant white-noise level in addition to the unavoidable Poisson fluctuations, while the \emph{alternative} model adds the Lorentzian to the same white-noise baseline. For a measured power $P$ in a given frequency bin and an expected model power $P_s$, the likelihood is given by \citep{1975ApJS...29..285G,2023Natur.613..253C}
\begin{equation}
\mathcal{L}(P\,|\,P_s)
\;=\;
e^{-(P+P_s)}
\sum_{m=0}^{\infty}\frac{(P\,P_s)^m}{(m!)^2},
\label{eq:lh}
\end{equation}
where the summation is truncated once the next term contributes less than $10^{-20}$ to the cumulative sum, following \citet{2019ApJ...871...95M}.

The observed PDS is fitted with the \emph{null} (constant) and the \emph{alternative} (QPO+constant) models using the likelihood function defined in Equation~\ref{eq:lh}. The fitting process is carried out using the nested sampling algorithm implemented in \texttt{MultiNest} \citep{2008MNRAS.384..449F, 2009MNRAS.398.1601F, 2014A&A...564A.125B, 2019OJAp....2E..10F}, which provides both posterior distributions of the model parameters and the Bayesian evidence for quantitative model comparison.

\begin{figure}
\centering
\includegraphics[width=0.9\linewidth]{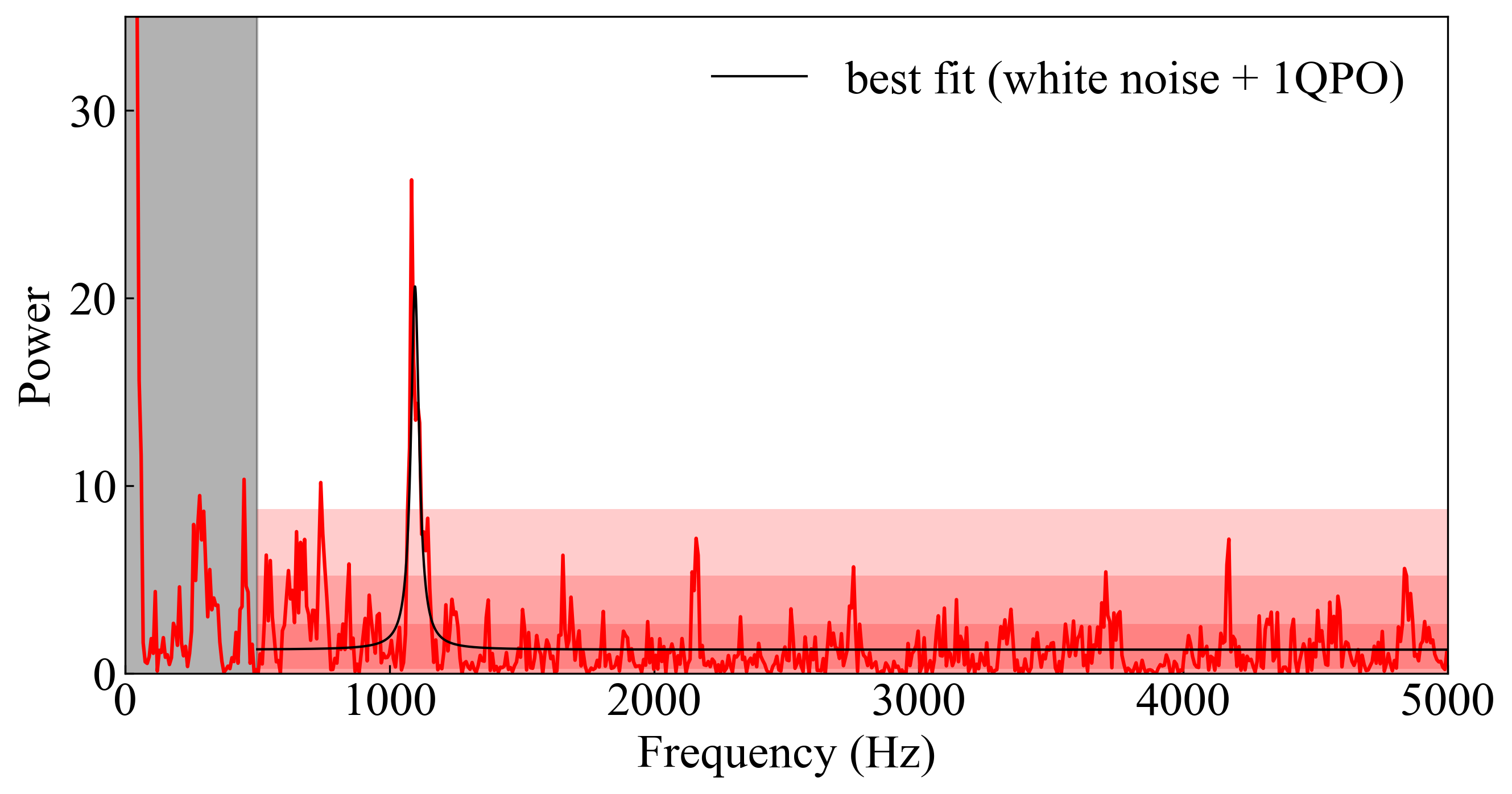}
\caption{
\noindent\textbf{PDS of the main peak of GRB~960616.}
The PDS is normalized following \citet{1975ApJS...29..285G}. The gray band marks the red-noise regime (0--500~Hz), which is excluded from the Bayesian model fitting. The red shaded regions indicate the 1$\sigma$, 2$\sigma$, and 3$\sigma$ confidence intervals expected for the white-noise component in addition to the pure Poisson noise (mean power $=1$; $\chi^2_2$ statistics). The black curve shows the best-fit model, consisting of a Lorentzian QPO component plus an additional constant white-noise term above the Poisson baseline. A significant peak is detected at $\sim$1083~Hz with a power of $\sim$26.3. Even when treating all 186,327 trials as independent, the probability of such a peak arising by chance remains only $\sim$$7\times10^{-7}$ (from the $\chi^2_2$ distribution), indicating that the periodic signal is very significant.}
\label{fig_psd}
\end{figure}

The best-fit QPO model accurately reproduces a narrow feature centered at about 1100~Hz, as shown in Figure~\ref{fig_psd}. The corresponding best-fit parameters are 
$A_{\rm QPO} = 19.36_{-3.64}^{+4.01}$, 
$A_{\rm white} = 0.28_{-0.05}^{+0.06}$, 
$f_0 = 1096.4_{-3.9}^{+3.8}$~Hz, and 
$\log_{10}\Delta f_0 = 1.27_{-0.08}^{+0.08}$. 
These values imply a quality factor of 
\[
Q \;\equiv\; \frac{f_0}{2\Delta f_0} \;\approx\; 29.4,
\]
which exceeds the Rayleigh coherence limit set by the main-burst duration. Specifically, since the main burst lasts only $\sim$24~ms, the Rayleigh frequency resolution is $\sim$42~Hz, implying a window-limited coherence of order $Q\!\sim\!45.7$ for a strictly main-burst measurement. By contrast, the fit over a longer time interval yields $Q\simeq29.4$, which is comparable in order but somewhat lower, consistent with the modest coherence dilution expected from using a longer time window and combining data from multiple detectors/energy channels. Taken together, the $\sim$1.1~kHz feature remains compatible with a coherent, nearly periodic modulation rather than a broad QPO. The Bayes factor (evidence ratio) in favor of the \emph{alternative} (QPO plus white-noise) model over the \emph{null} (white-noise-only) model is $10^{28.84}$, which is exceptionally high compared with the distribution of Bayes factors reported in \citet{2023Natur.613..253C}\footnote{The sample in \citet{2023Natur.613..253C} did not include segments restricted to BATSE channel~4; as seen in Figure~\ref{fig_wwz2}, the modulation is not significant in channel~3 and not visible in channels~1 and 2, so the signal would be missed by energy-aggregated or lower-energy selections.}.

This power-spectral evidence establishes a strong, narrow feature near 1.1~kHz with high coherence in the main emission episode. Based on this result, in the next subsection we directly fold the photon events at the candidate frequency to test the waveform morphology (sinusoidal vs.\ non-sinusoidal), and to measure the modulation amplitude with phase-resolved profiles.

\subsection{Epoch Folding and The Pulse Profile }

Motivated by the detected periodicity near 1100~Hz, we folded the photon events of GRB~960616 at this characteristic frequency to construct the phase-resolved pulse profile. The analysis was performed on the combined data from LAD0 and LAD1 within energy channel~4, using events in the main emission interval of $[0.095, 0.115]$~s. 

For a folding period $P = 1/1100$~s and a reference time $t_0 = 0$~s, the phase of each photon arriving at time $t_i$ (relative to $t_0$) was computed as \citep{1983A&A...128..245B}:
\begin{equation}
\phi_i = \frac{t_i}{P} - \left\lfloor \frac{t_i}{P} \right\rfloor,
\end{equation}
where $\lfloor \cdot \rfloor$ denotes the floor function, ensuring that $\phi_i \in [0, 1)$.

\begin{figure}
\centering
\includegraphics[width=0.9\linewidth]{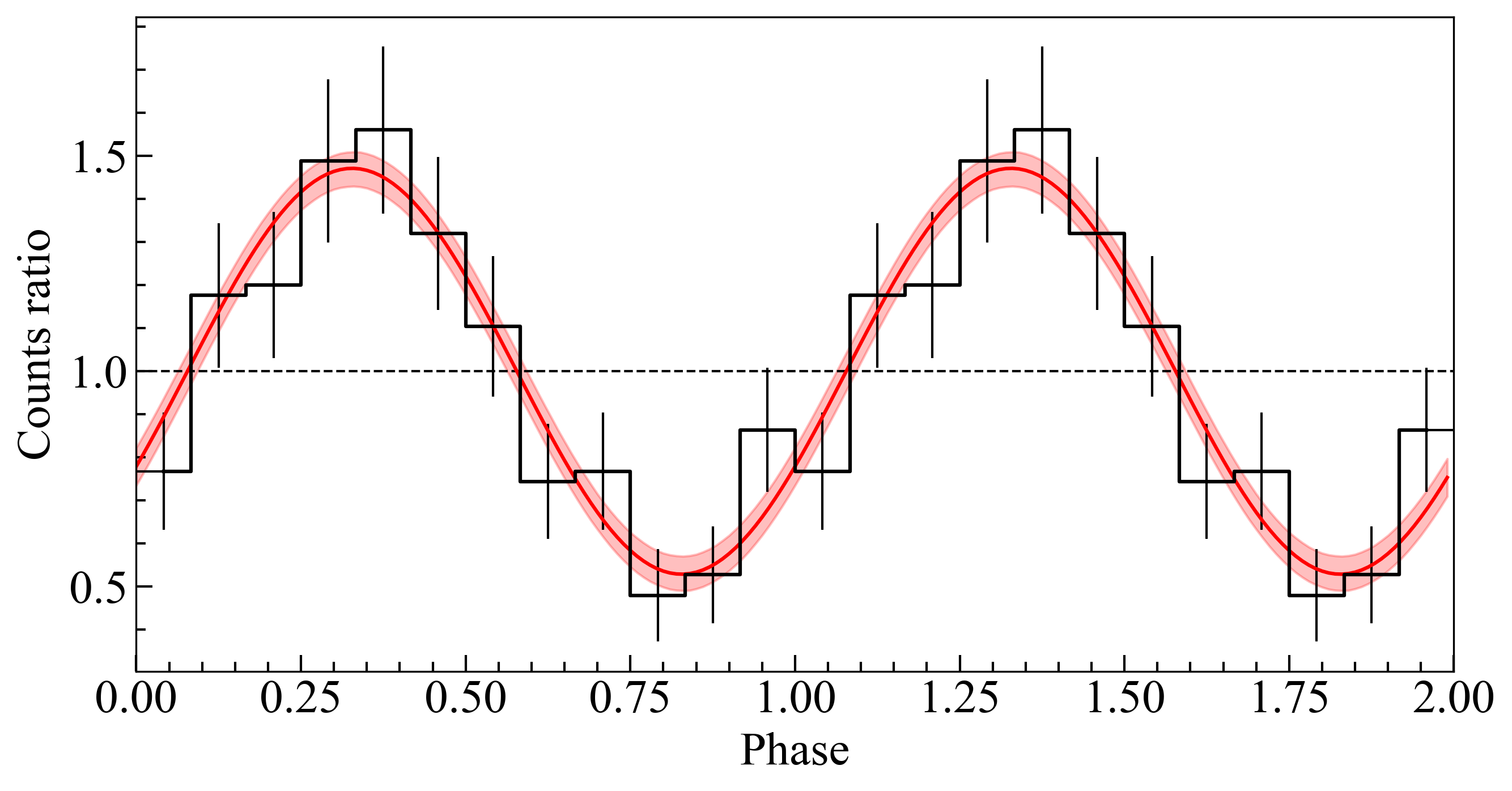}
\caption{
\noindent\textbf{Pulse profile of GRB~960616 folded at 1100~Hz.}
The black steps show the phase-folded pulse profile of GRB~960616, using data from LAD0 and LAD1 within energy channel~4. Error bars indicate the 1$\sigma$ uncertainties. 
The dashed horizontal line marks the mean count ratio (unity). 
The red curve represents the best-fit sinusoidal model, and the red shaded region denotes its 1$\sigma$ confidence interval.}
\label{fig_sine}
\end{figure}

The phases were divided into 12 equal bins over the interval $[0, 1)$, and repeated by one additional cycle for visualization. The count rates were normalized by the mean photon count so that the phase-averaged ratio equals unity. The folded profile {presented in Figure~\ref{fig_sine}} was then fitted with a sinusoidal model of the form
\begin{equation}
f(\phi) = A_{\rm s} \sin[2\pi(\phi + \phi_0)] + 1,
\end{equation}
where $A_{\rm s}$ is the fractional pulse amplitude relative to the mean count rate, and $\phi_0$ is the phase offset corresponding to the reference time $t_0$. A weighted least-squares fit yields best-fit parameters of $A_{\rm s} = 0.47 \pm 0.04$ and $\phi_0 = -0.08 \pm 0.02$ (1$\sigma$ uncertainties). The reduced $\chi^2 = 0.50$ ($p = 0.98$) indicates that the sinusoidal model provides an excellent fit to the phase profile, with residual scatter smaller than the nominal Poisson uncertainties. This high apparent goodness of fit likely reflects both the very high signal-to-noise ratio and the mild oversampling of the folded light curve, rather than model overfitting.

For comparison, the periodic signal reported in GRB~230307A exhibited a maximal root-mean-square{ (RMS)} fractional amplitude of $\sim$27\% \citep{2025NatAs...9.1701C}, corresponding to a pure-sine amplitude of $\sim$38\%. In contrast, the oscillation in GRB~960616 reaches a fractional amplitude of 47\%, signifying an unusually strong and coherent modulation of the prompt emission, which is consistent with a highly periodic origin.

{\subsection{Decomposition of the Light Curve}

Based on the PDS modeling and pulse-profile folding, the main emission of GRB~960616 exhibits a highly coherent oscillation with a well-defined period of $P=1/1100$~s. In the PDS, the preferred noise model requires an additional narrow Lorentzian component beyond white noise, and the best-fit quality factor indicates a coherent signal. Consistently, the epoch-folded pulse profile is well described by a near-sinusoidal modulation. Conditioned on the existence of a periodic signal at 1100~Hz, we decompose the light curve into a QPO component and a stochastic broadband component, enabling significance tests under realistic, data-driven noise realizations rather than idealized assumptions.

We first apply Seasonal--Trend decomposition using LOESS \citep[STL;][]{cleveland1990stl} to obtain a minimally-assumptive separation of the observed light curve $x(t)$ into a slowly varying trend, a periodic (seasonal) component, and a residual,
\begin{equation}
x(t)=T(t)+S(t)+R(t),
\end{equation}
where $T(t)$ is the trend, $S(t)$ captures variability at the prescribed period, and $R(t)$ contains the remaining aperiodic fluctuations. The decomposition is performed using the \texttt{STL} implementation in \texttt{statsmodels.tsa.seasonal} \citep{statsmodels2010}. The STL period is specified by the integer number of bins per cycle,
\begin{equation}
N_{\mathrm{per}}=\left\lfloor\frac{P}{\Delta t}\right\rfloor,
\end{equation}
with $\Delta t=256~\mu$s and $P=1/1100$~s. We apply the STL decomposition to the main emission interval (0.095--0.115~s), with the seasonal smoothing window of $4\times N_{\rm per}+1$ to preserve the QPO waveform, and a trend window of $2\times N_{\rm per}+1$ to capture the slowly varying baseline, yielding the trend $T(t)$, seasonal $S(t)$, and residual $R(t)$ components (Figure~\ref{fig_stl}).

To obtain a physically motivated refinement of this statistical decomposition, we model $S(t)$ and $R(t)$ using Gaussian-process (GP) regression with covariance kernels designed to represent (i) a coherent QPO and (ii) broadband red noise. In this framework, the GP kernel is the hypothesis about the underlying variability: given kernel hyperparameters, the GP specifies a multivariate normal likelihood for the data, with a covariance matrix set by the kernel plus the observational uncertainties \citep[e.g.,][]{2023ARA&A..61..329A}. We infer the posterior distributions of the kernel hyperparameters using MCMC sampling with the \texttt{emcee} ensemble sampler \citep{2013PASP..125..306F}, implemented with \texttt{celerite2} \citep{celerite1,celerite2}.

For the seasonal component, we adopt a one-term kernel consisting of a stochastically driven, damped harmonic oscillator (SHO) term plus an additional white-noise jitter term,
\begin{equation}
k_{\mathrm{seasonal}}(t,t')
=
k_{\mathrm{QPO}}(t,t')
+
\sigma_{\mathrm{jit}}^{2}\,\delta(t-t'),
\end{equation}
where $t$ and $t'$ are two time samples, $k(t,t')$ denotes the covariance between the process evaluated at those times, and $\delta(t-t')$ is the Dirac delta. For the residual component, we use a two-term kernel that allows any remaining quasi-periodic power to be distinguished from broadband stochastic variability,
\begin{equation}
k_{\mathrm{resid}}(t,t')
=
k_{\mathrm{QPO}}(t,t')
+
k_{\mathrm{red}}(t,t')
+
\sigma_{\mathrm{jit}}^{2}\,\delta(t-t').
\end{equation}
Here $k_{\mathrm{QPO}}$ is an SHO kernel whose characteristic frequency is fixed by the known QPO period $P$, while $k_{\mathrm{red}}$ is a low-$Q$ SHO kernel representing red-noise variability.

The one-sided power spectral density of an SHO process is
\begin{equation}
\mathcal{P}(\omega)=
\sqrt{\frac{2}{\pi}}\,
\frac{S_0\,\omega_0^4}{
\left(\omega^2-\omega_0^2\right)^2+\omega_0^2\omega^2/Q^2},
\end{equation}
where $\omega_0$ is the undamped natural angular frequency, $Q$ is the quality factor, and $S_0$ sets the normalization. In \texttt{celerite2}, the SHO kernel can equivalently be parameterized by an RMS amplitude $\sigma=\sqrt{S_0\,\omega_0\,Q}$, the quality factor $Q$, and the undamped oscillation period $\rho=2\pi/\omega_0$\footnote{See the \texttt{celerite} documentation for implementation details: \url{https://celerite.readthedocs.io}.}.

Accordingly, the GP fit to the seasonal component $S(t)$ samples three kernel hyperparameters, whereas the GP fit to the residual component $R(t)$ samples five hyperparameters, corresponding to the different kernel structures adopted for the two components. In practice, the posterior distributions are explored using the \texttt{emcee} ensemble sampler with 32 walkers for the seasonal fit and 40 walkers for the residual fit, each evolved for 3000 steps with the first 1500 steps discarded as burn-in. From the posterior samples, we reconstruct the posterior mean of the latent QPO process inferred from the seasonal component, $S_{\mathrm{QPO}}(t)$ (Figure~\ref{fig_stl}\textbf{b}), and the QPO and red-noise components extracted from the residual term, $R_{\mathrm{QPO}}(t)$ and $R_{\mathrm{red}}(t)$ (Figure~\ref{fig_stl}\textbf{c}). Together with the STL trend, these components define the intensity model
\begin{equation}
\lambda_{\mathrm{model}}(t)=
T(t)+S_{\mathrm{QPO}}(t)+R_{\mathrm{QPO}}(t)+R_{\mathrm{red}}(t),
\end{equation}
which we use for subsequent variability characterization and for generating noise-realistic Poisson realizations after incorporating the inferred jitter terms (Figure~\ref{fig_stl}\textbf{d}). The posterior constraints on the GP hyperparameters are summarized in Table~\ref{tab_gp}.

\begin{figure}
\centering
\includegraphics[width=0.9\linewidth]{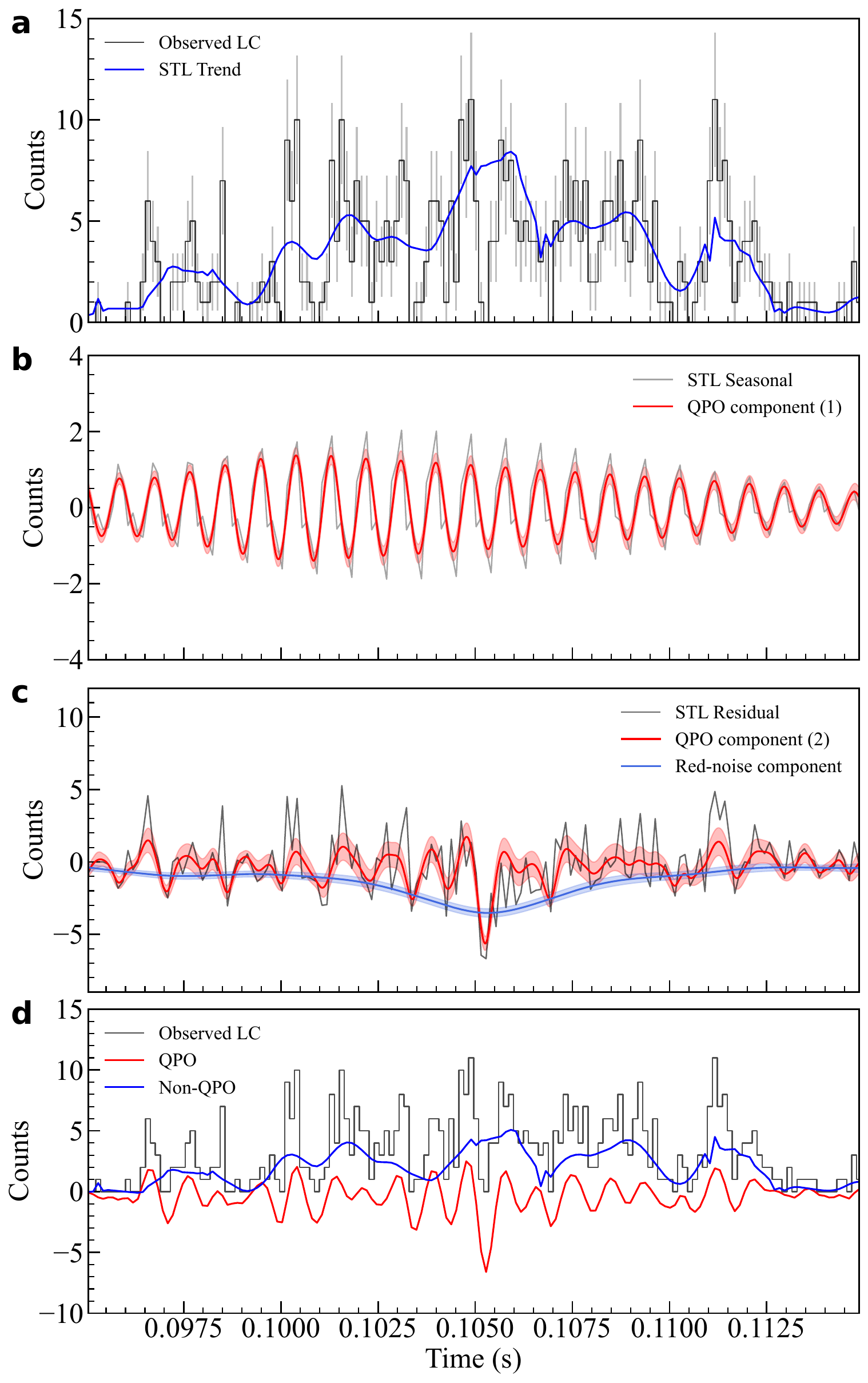}
\caption{
\noindent\textbf{Decomposition of the Light Curve of GRB~960616.}
\textbf{a:} Observed light curve in the main emission interval (0.095--0.115~s), together with the trend component extracted using STL. The trend traces the slowly varying baseline that provides the context for identifying rapid variability.
\textbf{b:} STL seasonal component and the GP reconstruction of the QPO signal. The red curve shows the GP-predicted QPO component, and the shaded region denotes the associated $1\sigma$ uncertainty.
\textbf{c:} STL residual component and its GP decomposition. The GP model separates the residual variability into a coherent QPO contribution (red) and a stochastic red-noise component (blue), both displayed as smoothed predictions with $1\sigma$ uncertainties.
\textbf{d:} Comparison between the observed light curve and the model reconstruction. The total model is the sum of the pure QPO signal (red; seasonal-QPO + residual-QPO) and the non-QPO component (blue; trend + red noise).}
\label{fig_stl}
\end{figure}

\begin{deluxetable*}{lccc}
\tablecaption{
Best-fit Gaussian-process parameters for the STL seasonal and residual components.
\label{tab_gp}}
\tablewidth{0pt}
\tablehead{
\colhead{Parameter} &
\colhead{Seasonal GP (QPO + jitter)} &
\colhead{Residual GP (QPO + red + jitter)} &
\colhead{Prior Range}
}
\startdata
$\sigma_{\rm qpo}$ 
& $0.718^{+0.405}_{-0.203}$
& $1.126^{+0.382}_{-0.367}$
& $[10^{-3}s_0,\; 10 s_0]$
\\
$Q_{\rm qpo}$
& $89.8^{+168.0}_{-49.9}$
& $1.523^{+0.969}_{-0.527}$
& $[1,\;10^3]$
\\
$\sigma_{\rm red}$
& --- 
& $0.899^{+0.471}_{-0.445}$
& $[10^{-3}s_0,\; 10 s_0]$
\\
$\tau_{\rm red}$ (s)
& ---
& $0.01298^{+0.00412}_{-0.00608}$
& $[P,\; 0.95\Delta t_{\rm span}]$
\\
$\sigma_{\rm jit}$
& $0.344^{+0.0522}_{-0.0449}$
& $0.548^{+0.334}_{-0.510}$
& $[10^{-3}s_0,\; 10 s_0]$
\\
\enddata
\tablecomments{
All quoted parameter values are posterior medians with $1\sigma$ uncertainties (16th--84th percentiles) evaluated in linear space. The quantity $s_0$ denotes the standard deviation of the decomposed light-curve segment used in each fit. Priors are taken to be uniform in \textit{logarithmic} space, that is, the MCMC sampling is performed in $\ln \sigma_{\rm qpo}$, $\ln Q_{\rm qpo}$, $\ln \sigma_{\rm red}$, $\ln \tau_{\rm red}$, and $\ln \sigma_{\rm jit}$. Here $P$ denotes the QPO period and $\Delta t_{\rm span}$ is the duration of the analyzed time interval.
}
\end{deluxetable*}

\subsection{Monte Carlo Tests Based on the GP Decomposition}

With the GP-based decomposition in hand, we proceed to test the statistical implications of the inferred quasi-periodic modulation under realistic, data-driven noise conditions. Adopting the strong prior established by the PDS and pulse-profile analyses, we explicitly separate all variability components associated with the 1100~Hz periodicity from those that are unrelated to it. The latter include the slowly varying trend, any additional broadband red-noise process, and excess white noise beyond Poisson counting statistics. As a result, the best-fit GP model provides a generative description of the GRB~960616 light curve that preserves the observed burst morphology while allowing the injection of an arbitrary QPO fractional RMS and mean photon count rate. This framework enables controlled Monte-Carlo simulations of light curves with the same duration as GRB~960616, but with user-specified QPO strength and counting statistics.

A first and primary question concerns the probability of false detections under the null hypothesis of no QPO, as well as the amount of radiative modulation implied if the QPO is real. To address this, we generate synthetic light curves with varying injected QPO fractional RMS and measure the Lomb--Scargle power \citep{2018ApJS..236...16V} at 1100~Hz for each realization\footnote{The Lomb--Scargle periodogram is computed using the \texttt{astropy.timeseries.LombScargle} implementation with the standard normalization; see \url{https://docs.astropy.org/en/stable/timeseries/lombscargle.html} for details.}. 

For each configuration, we perform 186,327 Monte-Carlo trials, matching the total number of independent searches conducted over all BATSE TTE data in this work. The resulting power distributions are shown in Figure~\ref{fig_sim}. In the absence of any injected QPO, the distribution of Lomb--Scargle power follows the expected exponential form for noise-dominated periodograms. Because the observed power far exceeds the maximum value obtained in all 186,327 simulations, we estimate its false-alarm probability by fitting the exponential tail of the distribution, yielding $p = 1.35\times10^{-9}$. After accounting for the $2000$ independent frequencies searched, this corresponds to a corrected probability of $2.69\times10^{-6}$, or a one-sided Gaussian significance of $\simeq4.55\sigma$. This independently confirms the robustness of the 1100~Hz signal.

As the injected QPO strength is increased, the simulated power distributions gradually transition from the central exponential form toward a non-central $\chi^2$ distribution, in agreement with theoretical expectations for a coherent signal embedded in noise (Figure~\ref{fig_sim}). From these simulations, we infer that the observed power is consistent with a QPO fractional RMS of approximately $40$--$50\%$, in good agreement with the value inferred from the GP posterior (Table~\ref{tab_gp}), for which the best-fit model yields a fractional RMS of $\simeq42.9\%$. We emphasize, however, that because the GP model imposes a strong periodic prior, the QPO contributions recovered in the seasonal and residual components should not be interpreted as physically independent. Both are likely manifestations of the same underlying modulation, and therefore the inferred QPO strength should be regarded as an upper limit. Given additional uncertainties in the GRB emission mechanism, we conservatively estimate that the QPO modulates no more than $\sim8\%$ of the total radiated energy in GRB~960616.

A second test addresses the unusually high photon count rate of GRB~960616 relative to the broader BATSE short-GRB population (Section~\ref{sec:a3}). Fixing the QPO fractional RMS to $42.9\%$, we repeat the simulations for a range of mean count rates. For reference, GRB~960616 has an average count rate of $\langle\lambda\rangle\simeq3.205$ in the 0.095--0.115~s interval, using 256~$\mu$s bins in channel~4 and combining LAD~0 and LAD~1. As shown in Figure~\ref{fig_sim}, even at a mean count rate as low as $\langle\lambda\rangle = 0.5$ counts per 256~$\mu$s bin, the 186,327 Monte Carlo realizations still produce at least one light curve with a Lomb--Scargle power comparable to the observed value. The explored range of mean count rates, $\langle\lambda\rangle = 0.5$--$3.5$, is defined within the same temporal resolution and energy selection used for GRB~960616. Although this quantity is not directly comparable to the 16~ms peak rates shown in Figure~\ref{fig_a3}, the lower end of this range corresponds to light curves that are substantially fainter than GRB~960616 when expressed in the same instrumental and temporal sampling. This demonstrates that, given a similar QPO modulation strength, bursts significantly fainter than GRB~960616 would still be expected to yield detectable QPO signatures. The fact that GRB~960616 remains unique therefore suggests that the physical conditions required to generate such strong modulation are intrinsically rare among short GRBs\footnote{We note that these simulations assume comparable signal-to-noise ratios and do not explicitly separate background contributions, rendering the test conservative.}.

Finally, simulations based on the non-QPO GP model provide insight into the presence of marginally significant features at other frequencies. As illustrated in Figures~\ref{fig_wwz1} and~\ref{fig_wwz2}, short-GRB light curves with sufficient signal-to-noise can naturally produce apparent excesses at high frequencies over 500~Hz. Even after removing the modeled QPO component, broadband variability intrinsic to the burst can leak power into these frequencies (Section~\ref{sec:a4}). This demonstrates that such marginal features do not necessarily indicate additional physical oscillation components, and highlights the importance of targeted, hypothesis-driven tests when assessing the significance of high-frequency variability in short GRBs.

\begin{figure}
\centering
\includegraphics[width=\linewidth]{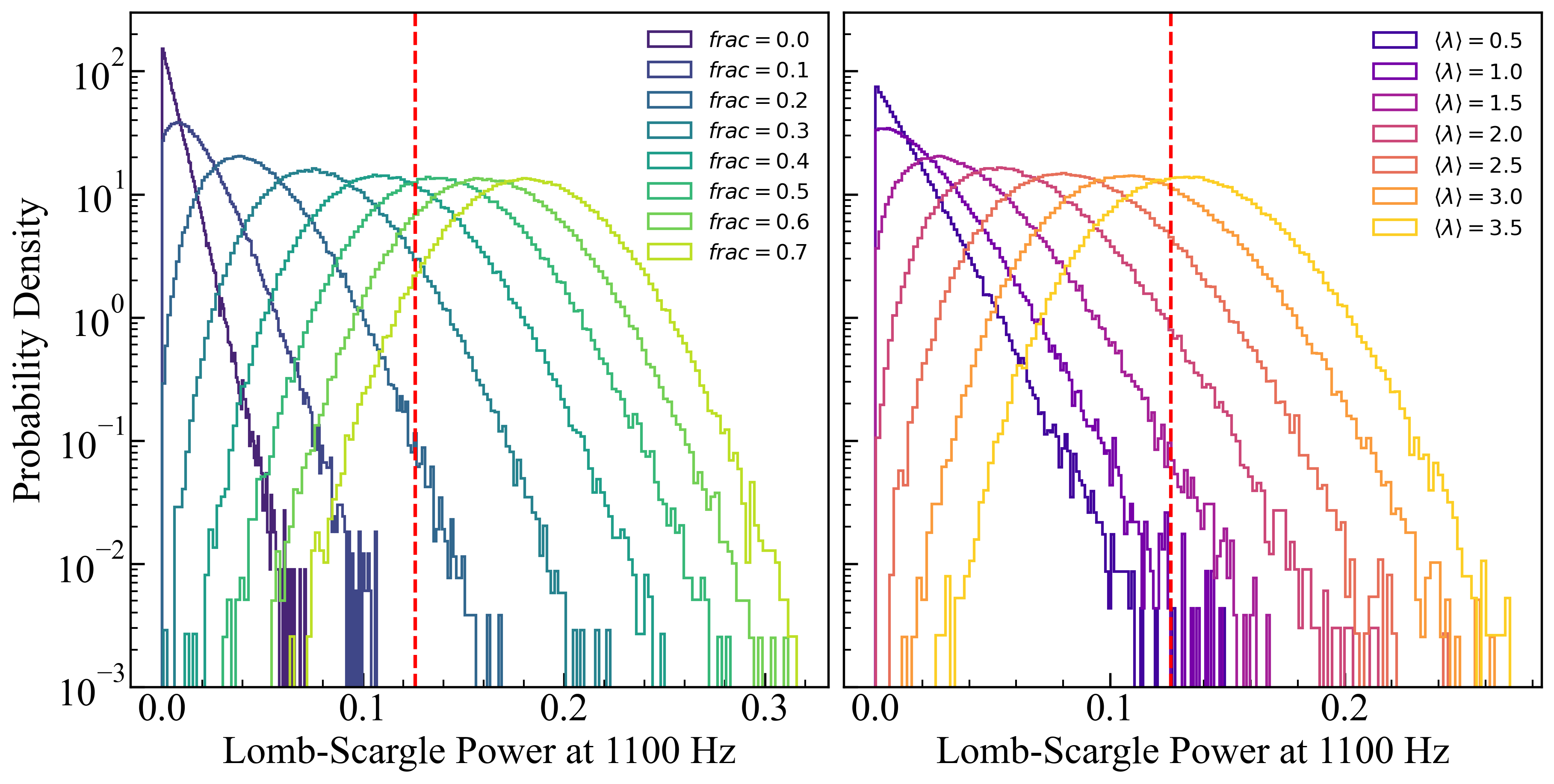}
\caption{\textbf{Distributions of Lomb--Scargle power at 1100~Hz for simulated light-curve realizations.} The simulations are based on the GP decomposition of GRB~960616, with controlled variations in either the intrinsic QPO fractional RMS (left panels, $frac$) or the mean count rate per 256-$\mu$s bin (right panels, $\langle\lambda\rangle$), while all other model components are held fixed. Each distribution is constructed from Monte Carlo realizations of Poisson-sampled light curves. The vertical dashed line indicates the Lomb--Scargle power measured from the observed data.}
\label{fig_sim}
\end{figure}

}

\section{Summary and Discussion}\label{sec:sum}
The broad energy coverage and high temporal resolution of BATSE, together with its high photon statistics enabled by the large-area detector modules, make it particularly suitable for detecting millisecond (quasi-)periodic signals in prompt gamma-ray emission. Motivated by the recent detection of kilohertz periodicity in GRB~230307A \citep{2025NatAs...9.1701C}, we systematically searched for coherent oscillations in 532 short GRBs using their TTE data. After excluding two events dominated by extreme red noise contamination, we identified a single burst, GRB~960616 (trigger~5502), exhibiting a highly significant periodic signal. This periodicity is centered at $\sim$1100~Hz, persists throughout the main emission episode, and shows the strongest power in the highest BATSE energy channel ($\gtrsim$320~keV), with a fractional amplitude reaching $47\%\pm4\%$.

It is worth emphasizing that the methodology employed in this work differs substantially from the approach used to identify the kilohertz QPO in GRB~910711 by \citet{2023Natur.613..253C}. That study was motivated by the anticipation of a different type of central engine and focused on a transient oscillatory feature whose power spectrum follows a Lorentzian-like, quasi-continuous distribution---that is, a series of broadband peaks around a centroid frequency. Such a Lorentzian profile naturally arises from the Fourier transform of a damped oscillatory process and therefore provides a physically motivated template for quasi-periodic signals. In this framework, the statistical assessment relies on a template-matching approach in the frequency domain, enabling Bayesian evidence comparisons between models with and without a Lorentzian component. In contrast, the present method does not assume any underlying oscillatory form. Instead, the search targets statistically exceptional peaks that stand out from the ensemble of maximal powers across all frequency trials, following the framework of extreme-value statistics. As a result, while a relatively broad Lorentzian feature with modest peak power (see, e.g., Figure~2 of \citealt{2023Natur.613..253C}) may appear significant in a Bayesian template comparison, it would not emerge as an outlier in an extreme-value distribution, which inherently prioritizes narrow, high-coherence, and statistically rare peaks.

The general properties of the detected signal in GRB~960616 are consistent with the theoretical framework and observational conditions proposed for GRB~230307A \citep{2025NatAs...9.1701C}, which invoke a magnetar central engine powering a highly magnetized outflow that dissipates through the internal-collision-induced magnetic reconnection and turbulence (ICMART) process \citep{2011ApJ...726...90Z}. Specifically: 
\begin{enumerate}
\item The signal in GRB~960616 is sharply peaked in the frequency domain, consistent with a nearly periodic modulation. Its period of $\sim$0.9~ms matches the expected spin timescale of a newly formed millisecond magnetar and cannot plausibly arise from precessional variations of the jet, which generally produce broader and lower-frequency features \citep[e.g.,][]{2010A&A...516A..16L}.
{In addition, variability originating in the emission region itself is also unlikely to sustain such a millisecond-level coherence. Dissipative processes such as magnetic reconnection \citep[e.g.,][]{2011ApJ...726...90Z} or internal shocks \citep[e.g.,][]{2004RvMP...76.1143P} are inherently stochastic and typically lack a stable phase reference, leading to rapidly decorrelating or broadband variability rather than a sharply defined periodic signal. The detection of a coherent modulation persisting throughout the entire prompt-emission episode therefore strongly favors an origin tied to a central-engine clock, consistent with a millisecond magnetar interpretation.}
\item The modulation in GRB~960616 persists throughout the entire duration of the prompt emission. 
{This difference in duty cycle contrasts with GRB~230307A, where the $\sim$160~ms periodic episode occupies only a tiny fraction of the minute-long prompt emission \citep{2025NatAs...9.1701C}, while in GRB~960616 the modulation spans a substantial fraction of the entire duration. Nevertheless, both cases remain compatible within the same ICMART model. In this framework, the prompt emission is powered by energy dissipation in multiple reconnection-driven mini-jets embedded within a globally structured (e.g., striped) magnetic field. While the mini-jets themselves provide the sites of radiative energy release, any underlying periodicity is more naturally associated with the large-scale magnetic or central-engine structure. The observability of such a periodic modulation therefore depends on whether the emission from the mini-jets within the relativistic beaming cone departs sufficiently from statistical symmetry. Specifically,} 
the orientations and spatial distribution of these mini-jets are expected to be random and approximately symmetric, such that their combined emission normally averages out any spin-induced modulation. However, when the number of effectively radiating mini-jets becomes small, this statistical symmetry can be broken, allowing the underlying periodic modulation to emerge in the observed emission. 
{In relativistic outflows, Doppler beaming always determines that the observed emission is dominated by mini-jets whose velocities are oriented close to the line of sight. During the main prompt phase, however, a large number of simultaneously radiating mini-jets typically contribute within the relativistic beaming cone, so that their combined emission is approximately symmetric and any underlying periodic modulation is averaged out. In GRB~230307A, the periodic signal was detected only within a brief temporal window, which can be understood as a situation in which this statistical averaging temporarily failed. As the burst entered the high-latitude arrival phase, light-travel-time effects limited the number of mini-jets contributing at a given observer time \citep{2025NatAs...9.1701C}. In this regime, the beaming-induced selection effect---normally masked by angular averaging---became prominent, such that the observed emission was dominated by only a few favorably oriented mini-jets. The resulting breakdown of statistical symmetry allowed the underlying periodic modulation to emerge. By contrast, GRB~960616 likely represents a different regime in which the number of simultaneously radiating mini-jets was intrinsically small throughout the burst. In this case, the emission within the beaming cone never reached full statistical symmetry, preventing complete averaging and allowing the spin-induced periodicity to persist over essentially the entire prompt emission episode.}
\item {The modulation is more pronounced at higher photon energies and becomes progressively weaker toward lower energies. This energy dependence follows naturally from the same Doppler-beaming and mini-jet asymmetry arguments discussed above. In the ICMART framework, relativistic beaming always favors emission from mini-jets whose motions are oriented closer to the line of sight. High-energy photons are therefore dominated by a small subset of favorably aligned mini-jets, whereas lower-energy photons receive substantial contributions not only from these on-axis emitters but also from a much larger population of more obliquely oriented mini-jets whose emission is Doppler-shifted to lower energies. As a result, the effective number of contributing emitters is intrinsically larger at low energies, leading to stronger statistical averaging and a reduced visibility of any underlying periodic modulation. Conversely, at higher energies, where fewer mini-jets dominate the emission, the asymmetry is enhanced and the periodic signal becomes more pronounced. Although the BATSE TTE data do not permit finer energy subdivision, the trend observed in Figure~\ref{fig_wwz2} is consistent with these expectations and in qualitative agreement the energy-dependent behavior reported for GRB~230307A \citep{2025NatAs...9.1701C}. More generally, in the presence of a coherent periodicity, such an energy dependence is difficult to account for without invoking Doppler beaming and mini-jet statistics: in more conventional emission scenarios without pronounced directional substructure \citep[e.g., internal shocks,][]{2004RvMP...76.1143P}, photons at different energies are generally produced by the same macroscopic emitting region and are subject to similar temporal averaging. While relativistic motion can Doppler-shift the photon energies depending on the observer’s angle, it does not naturally reduce the number of independent emitting elements contributing at a given time. As a result, there is no obvious mechanism to selectively preserve a coherent millisecond modulation at high energies while washing it out at lower energies.}
\end{enumerate}

Alternatively, one might consider an origin related to magnetar giant flares (MGFs), which are also capable of producing high-frequency QPOs \citep{2021Natur.600..621C}. However, GRB~960616 shows several features that are inconsistent with the properties of known extragalactic MGFs. First, it exhibits a distinct soft precursor separated from the main burst by only $\sim$10~ms---significantly shorter than the $\sim$100-second separation observed in the MGF precursor from SGR 1806-20 \citep{2005Natur.434.1098H}. Second, although GRB~960616 exhibits a bright main burst with rapid variability, its emission shows no evidence of the extended, slowly decaying tail (typically lasting $\sim$100~ms) that characterizes bright extragalactic MGFs \citep[e.g., GRB~200415A;][]{2020ApJ...899..106Y,2021Natur.589..207R,2021Natur.589..211S}. However, because no significant signal can be extracted from the STTE data recorded by SDs for GRB~960616, and the other available spectral data types have coarse temporal resolution ($\gtrsim$1~s), it is not possible to extract a reliable time-integrated spectrum to determine its spectral characteristics. In addition, the $\sim$2$^{\circ}$ localization uncertainty given by BATSE LADs prevents a secure identification of its host galaxy. Therefore, although GRB~960616 exhibits several characteristics that differ from known MGFs, we cannot entirely exclude such an origin. Nevertheless, given all currently available evidence, its properties are far more consistent with those of a typical merger-type short GRB.

The implications of this detection are twofold, encompassing both observational and theoretical perspectives. From the observational standpoint, the discovery of a second stable millisecond-period signal---this time modulating the entire main emission episode of GRB~960616---strongly suggests that, at least in some cases, the rotational signature of a GRB central engine can survive propagation through the relativistic jet and imprint itself onto the prompt emission. According to the framework proposed for GRB~230307A \citep{2025NatAs...9.1701C}, the visibility of such a modulation likely requires additional conditions, including the formation of a rapidly rotating, long-lived millisecond magnetar following the merger, a magnetically dominated dissipation process and sufficient photon statistics at high energies. These requirements may explain why GRB~960616 stands out as the only case among the 532 analyzed short GRBs to exhibit a clear periodic signature. 
{Rather than reflecting limitations of the search, this rarity likely points to the restrictive physical conditions required for such signals to become observable. More broadly, our results help inform analogous searches to other GRB populations, particularly long GRBs, where longer durations and higher photon statistics may offer additional opportunities to reveal transient millisecond periodicities under favorable conditions.}

From the theoretical perspective, the persistence of a highly coherent $\sim$1~kHz modulation through the jet dissipation region presents a significant challenge to current models of GRB emission. While the striped-wind configuration proposed in \citet{2025NatAs...9.1701C} can, in principle, transmit the rotational imprint of a newborn millisecond magnetar into the emission region, it remains unclear whether such a highly ordered magnetic structure can survive the extensive magnetic reconnection and turbulence during the dissipation process. In particular, the striped geometry is intrinsically favorable for magnetic reconnection, which may lead to extremely rapid energy release and magnetic-field dissipation before the GRB jet reaches the dissipation radius \citep{2025ARA&A..63..127S}. Although this process could be modified under extreme relativistic conditions, maintaining large-scale magnetic coherence throughout the jet remains a major theoretical challenge. Within the existing GRB framework and without invoking additional new physics or alternative radiation mechanisms, understanding how such a coherent, millisecond-scale modulation can be stably transported from the central engine to the emission region likely requires global magnetohydrodynamic simulations that self-consistently link the central engine, jet launching, and magnetic dissipation. Such simulations will be crucial for testing whether the observed periodic modulation can persist under realistic GRB conditions, despite their considerable computational demands.

%% Please use the acknowledgment and contribution environments. This will 
%% be anonomyized when the "anonymous" style option is used. 
\begin{acknowledgments}
We acknowledge the support by the National Natural Science Foundation of China (grant Nos. 12573046, 12121003 and 13001106), the National Key Research and Development Programs of China (2022YFF0711404, 2022SKA0130102), the National SKA Program of China (2022SKA0130100), the science research grants from the China Manned Space Project with NO. CMS-CSST-2021-B11, the Fundamental Research Funds for the Central Universities, and the Program for Innovative Talents and Entrepreneurs in Jiangsu.

\end{acknowledgments}

% \begin{contribution}
% contribution
% \end{contribution}

% \software{astropy \citep{2013A&A...558A..33A,2018AJ....156..123A,2022ApJ...935..167A}}

\bibliography{ms}{}
\bibliographystyle{aasjournalv7}

\appendix
\restartappendixnumbering

\section{Example for data reduction}\label{sec:a1}
Here, trigger~575 (corresponding to GRB~910725) is taken as an example to illustrate how 100~ms event segments are extracted from the full TTE record using the Bayesian blocks method.

\begin{figure}[h!]
\centering
\includegraphics[width=0.9\linewidth]{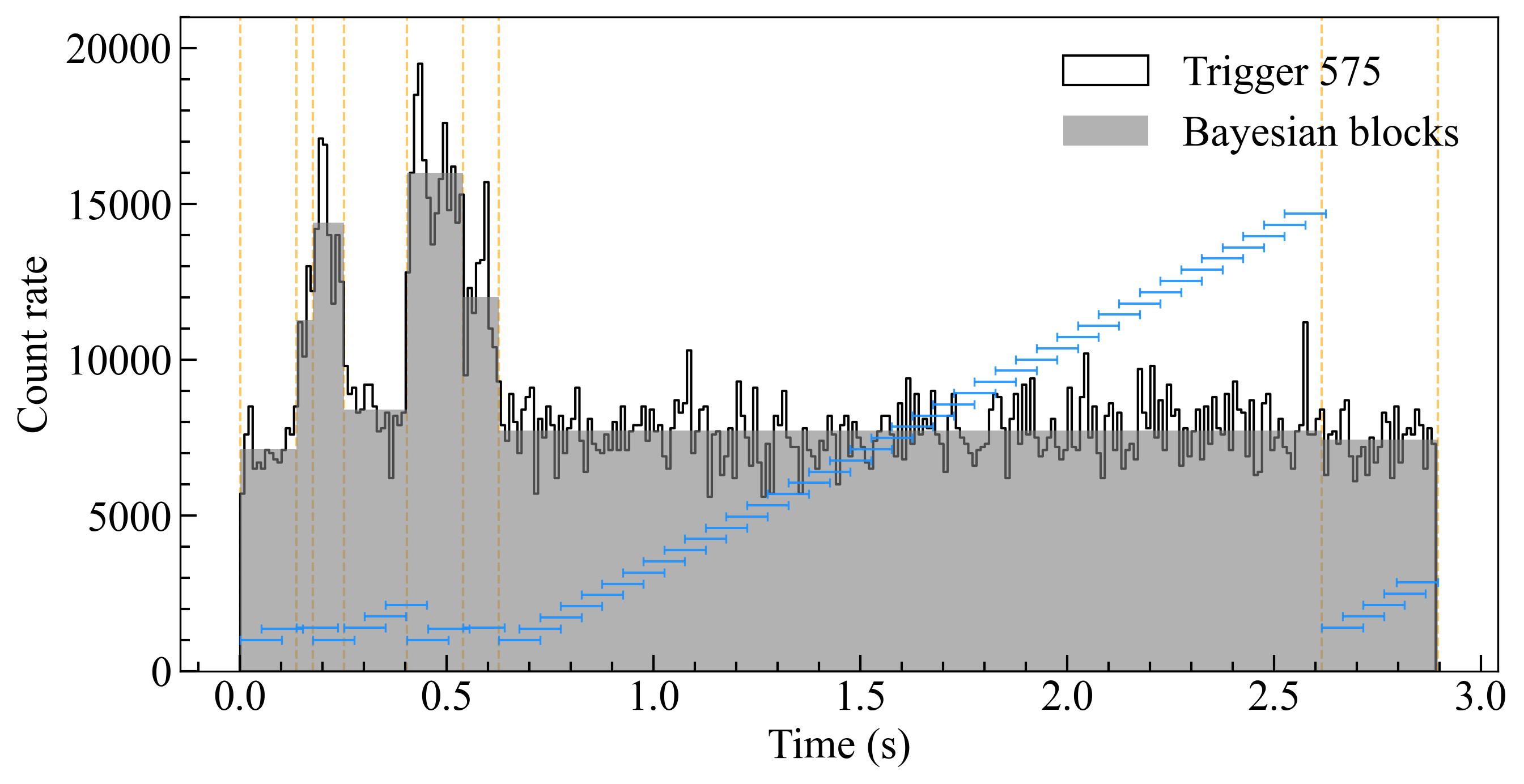}
\caption{
\noindent\textbf{Example of Bayesian-block-based segmentation and event extraction.}
Bayesian blocks are computed from the combined data of all triggered LADs across all energy channels. 
Orange dashed lines mark the boundaries of individual blocks, while blue horizontal error bars represent the 100~ms segments used to extract event data from each LAD in each energy channel. 
For each block, segmentation begins at the left boundary and continues until the full duration of the trigger's TTE data is covered. If a block is longer than 100~ms, consecutive segments overlap by 50\%.}
\label{fig_a1}
\end{figure}

\clearpage

\section{Two spurious periodic signal}\label{sec:a2}

\begin{figure}[h!]
\centering
\includegraphics[width=0.9\linewidth]{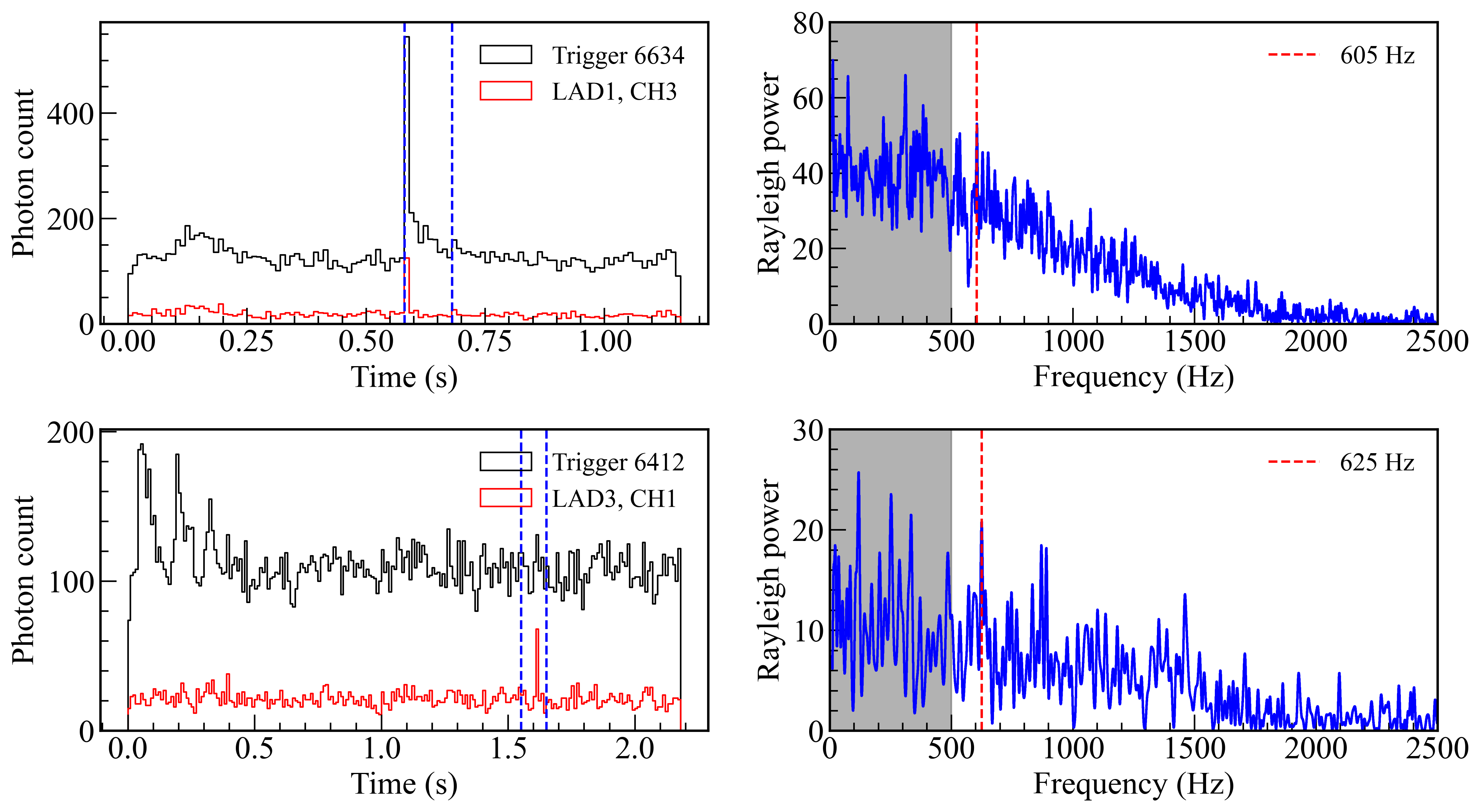}
\caption{
\noindent\textbf{Examples of spurious periodic signals.} Light curves and corresponding periodograms are shown for two triggers that exhibit apparent but non-physical periodicities. 
The upper panels show trigger~6634: the light curve (left) and the periodogram (right) of the 100~ms interval marked in the left panel, calculated for LAD3 in energy channel~1. The lower panels show trigger~6412: the light curve (left) and the periodogram (right) of the marked 100~ms interval, calculated for LAD1 in energy channel~3. 
Both cases display pronounced red-noise excesses extending into the kilohertz range, whereas the mean power of a purely Poisson noise process should be unity. 
The extreme red noise in trigger~6634 originates from a brief, intense spike within the burst, consistently observed across all detectors and energy bands. In contrast, the excess in trigger~6412 arises from an isolated deviation preceding the main burst, detected only in one low-energy channel of a single detector, suggesting that it is more likely due to a transient particle event rather than genuine GRB emission.}
\label{fig_a2}
\end{figure}

\clearpage

\section{Distribution of peak count rates of 532 BATSE short GRBs}\label{sec:a3}

\begin{figure}[h!]
\centering
\includegraphics[width=0.9\linewidth]{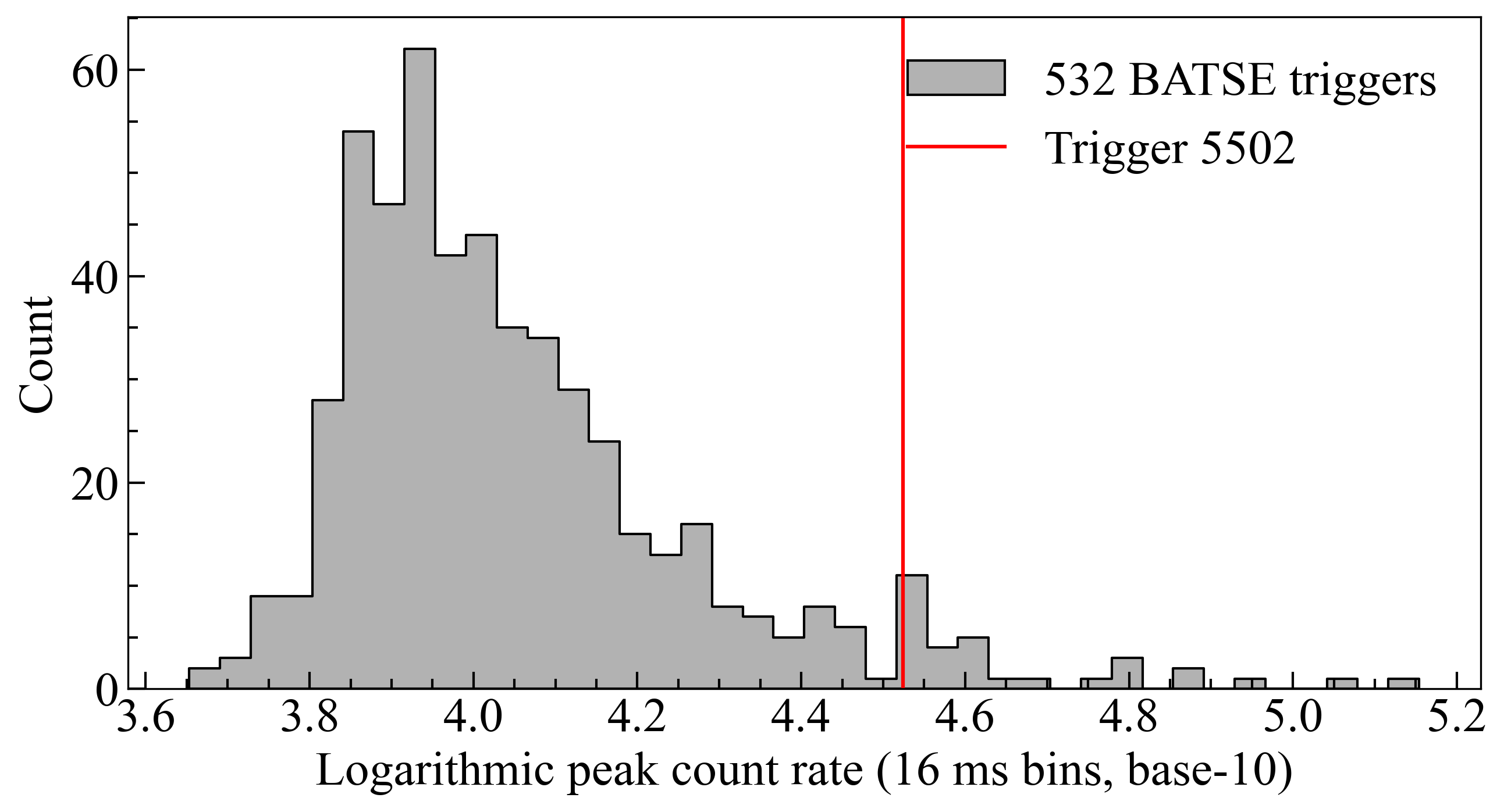}
\caption{\noindent\textbf{Distribution of peak count rates for 532 BATSE short GRBs.}
For each trigger, the peak count rate is measured from the brightest detector using 16~ms binned light curves. The four energy channels of each detector are combined, and the light curves are not background-subtracted. 
The position of trigger~5502 (GRB~960616) is marked by the red vertical line, showing that this burst lies near the upper end of the overall distribution.}
\label{fig_a3}
\end{figure}

\clearpage

{

\section{Periodogram distributions from simulated light curves without the QPO}\label{sec:a4}

Figure~\ref{fig_a4} shows an ensemble of 1000 periodograms computed from simulated light curves in which the QPO component has been removed. Even under the no-QPO hypothesis, the simulations frequently produce apparent peaks in the 500--1000~Hz range with Lomb--Scargle powers exceeding the observed value. This is expected for short-duration, Poisson-sampled bursts with strong broadband (red-noise-like) variability, where scanning many frequencies naturally yields occasional high-power excursions from the extreme-value tail of the noise distribution. These excesses explain why numerous seemingly significant peaks seen in Figures~\ref{fig_wwz1} and~\ref{fig_wwz2} can appear across a broad frequency range, particularly since a coherent 1100~Hz modulation can naturally redistribute power to neighboring frequencies and integer subharmonics under realistic observational conditions.

As for the higher-frequency features appearing in the WWZ spectrogram (e.g., the $\sim$2000~Hz excess visible in the lower-left panel of Figure~\ref{fig_wwz2}), these peaks are not produced by the underlying light curve itself in the absence of a QPO, but can arise naturally from the WWZ calculation under the adopted time--frequency resolution. In particular, the $\sim$2000~Hz feature in Figure~\ref{fig_wwz2} appears as a localized enhancement lasting $\sim$5--10~ms. For our chosen WWZ decay constant, $c = 1/(1100~\mathrm{Hz} \times 0.1~\mathrm{s})^2$, the corresponding effective temporal scale at frequency $f$ can be estimated as
\begin{equation}
\Delta t_{\rm eff}(f) \sim \frac{1}{2\pi f \sqrt{c}},
\end{equation}
which yields $\Delta t_{\rm eff} \sim 9$~ms at $f \sim 2000$~Hz. Consequently, the observed enhancement occupies only a single effective WWZ time window and does not persist across adjacent windows. Such isolated features are a natural and expected consequence of the WWZ time--frequency localization applied to short-duration light curves, and do not constitute credible evidence for a coherent periodic signal.

\begin{figure}[h!]
\centering
\includegraphics[width=0.5\linewidth]{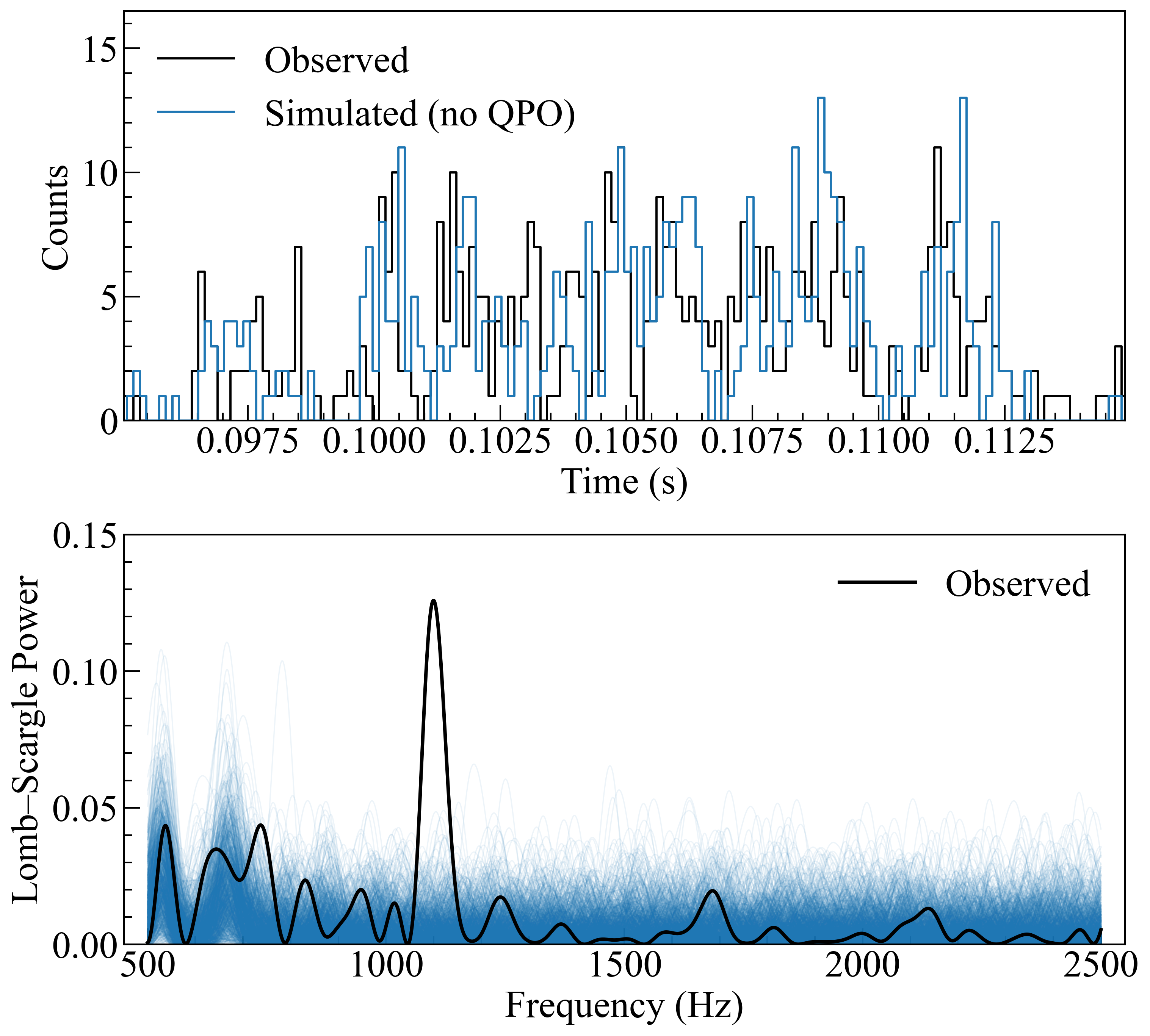}
\caption{\noindent\textbf{Lomb--Scargle periodogram distributions from GP-based simulations without a QPO.}
Top panel: comparison between the observed light curve of GRB~960616 and one representative simulated realization generated under the null hypothesis of no QPO, using the GP-based non-QPO model.
Bottom panel: Lomb--Scargle periodogram of the observed light curve (black curve) compared with periodograms from 1000 simulated light curves without a QPO (blue curves).
}
\label{fig_a4}
\end{figure}

}

\end{document}